\documentclass[twocolumn, prx, superscriptaddress, notitlepage]{revtex4-2}
\usepackage{graphicx}
\usepackage{physics}
\usepackage{float} 
\usepackage{subfigure} 
\usepackage{color}
\usepackage{amsmath}
\usepackage{amsfonts}
\usepackage[normalem]{ulem}
\usepackage{xspace}
\usepackage[dvipsnames]{xcolor}
\usepackage{dcolumn}
\usepackage[breaklinks,colorlinks,linkcolor=blue,citecolor=blue,urlcolor=blue]{hyperref}
\usepackage{lineno}

\begin{document}
\title{Resonant dynamics of 1D dipole-conserving Bose-Hubbard model with time-dependent tensor electric fields}

\author{Jiali Zhang}
\affiliation{Center of Intelligence and Quantum Science (CIQS), International Joint Laboratory on Quantum Sensing and Quantum Metrology, School of Physics, Huazhong University of Science and Technology, Wuhan 430074, China}

\author{Shaoliang Zhang}
\email{shaoliang@hust.edu.cn}
\affiliation{Center of Intelligence and Quantum Science (CIQS), International Joint Laboratory on Quantum Sensing and Quantum Metrology, School of Physics, Huazhong University of Science and Technology, Wuhan 430074, China}
\affiliation{Hubei Key Laboratory of Gravitation and Quantum Physics, Institute for Quantum Science and Engineering, Huazhong University of Science and Technology, Wuhan 430074, China}
\affiliation{Hefei National Laboratory, Hefei 230088, China}

\date{\today}

\begin{abstract}
Recently, tensor gauge fields and their coupling to fracton phases of matter have attracted more and more research interest, and a series of novel quantum phenomena arising from the coupling has been predicted.
Here, we propose a tunable platform using a one-dimensional dipole-conserving Bose–Hubbard chain subject to a periodically driven quadratic potential, which realizes a time-dependent rank-2 electric field. By tuning the drive frequency into resonance with the on-site interaction, the dynamics of dipoles and fractons change drastically from constrained motion to near-ballistic expansion via absorption of a quanta from the tensor electric fields, while the drive amplitude provides a knob for control. We identify experimentally accessible observables: wave packet radius of dipoles and fractons, which quantifies the speed of expansion, and state populations. Our work provides a possible approach to manipulate the dynamics of dipole-conserving quantum systems via tensor gauge fields.
\end{abstract}
\maketitle 

\section{Introduction}
The interplay between matter and gauge fields has provided important tools for exploring novel quantum phenomena~\cite{thouless1982,tsui1982,stormer1999fractional,konig2007}. Beyond conventional gauge fields that satisfy vector gauge invariance, such as electromagnetic fields, research of tensor gauge fields that obey higher-rank tensor gauge invariance and their coupling with matter fields has generated significant interest recently~\cite{vijay2015, vijay2016,prem2018}. 
The matter fields coupled with tensor gauge fields are called fracton phase of matter~\cite{nandkishore2019,pretko2020,ma2018fracton}. Unlike conventional excitations, the movement of an individual fracton is prohibited, and only the combination of a few fractons, such as a dipole 
formed by a particle-hole pair, can move along the constrained directions~\cite{pretko1,pretko2,boesl2024}. The restricted mobility of the fracton arises from the conservation of the dipole or multipole moments~\cite{pretko2018fracton,pretko2018fracton1} in systems that follow the tensor gauge theory. Therefore, tensor gauge fields can be exploited to control fracton phases and their dynamics, thereby linking these systems to glassy quantum dynamics~\cite{prem2017glassy}, elasticity theory~\cite{pretko2018fracton}, and gravitational fields, while also having broad potential applications in fields such as quantum information storage~\cite{haah2011local} and quantum computation~\cite{ma2017fracton,pai2019localization}.

Despite extensive theoretical progress on fracton phases and their coupling to tensor gauge fields, fundamental challenges remain unresolved, such as the creation and manipulation of individual fracton excitations in natural materials and, crucially, the direct probing of their coupling to rank-2 gauge fields.Beyond searching for fractons phases in natural materials, artificial quantum systems such as ultracold atomic lattices and superconducting quantum circuits provide powerful platforms for investigating fracton phases and their coupling to synthetic tensor gauge fields~\cite{zhang2025synthetic,myerson2022construction}. Recently, the dipole-conserving Bose-Hubbard model with correlated tunneling has been realized~\cite{yang2020,zhou2022,scherg2021,kohlert2023}. In this model, isolated particle or hole excitations can be considered as fractons, and their constrained
dynamics have been experimentally observed~\cite{adler2024,will2024,moudgalya2022quantum,sala2020ergodicity,khemani2020localization,Pai}. Moreover, dipole formed by a particle-hole pair can move along the constrained directions. So new phases of matter, such as dipolar Bose-Einstein condensate and dipolar Mott insulators, as well as the dynamics of dipoles and fractons, have been predicted~\cite{lake2022,lake2023,xu2024,Zechmann2023,Zechmann2024}. Motivated by the absence of coupling to tensor gauge fields in these works, our recent work~\cite{zhang2025synthetic} proposes a scheme to construct a synthetic rank-2 electric field, which can be utilized to drive dipole excitations. Although novel phenomena, such as dipolar Bloch oscillation, have been predicted, the controlled driving of an isolated fracton remains to be explored. Because the movement of an individual fracton is always accompanied by the creation of additional dipoles, an extra energy cost is necessary. Consequently, it is highly desirable to explore methods for simulating time-dependent tensor gauge fields and for engineering the dynamics of fractons.

In this work, we propose a scheme to simulate a time-dependent rank-2 tensor electric field in a one-dimensional dipole-conserving Bose-Hubbard model by introducing a periodically driven quadratic potential, thereby enabling investigations of the dynamics of fractons and dipoles under time-dependent tensor gauge fields. Unlike the static tensor electric field, which cannot drive an individual fracton, we show that the time-dependent tensor electric field drastically enhances the mobility of the fractons when its drive frequency is resonant with the on-site interaction,
as the movement of the fracton is accompanied by the absorption of drive quanta. Furthermore, the tensor electric field with resonant frequency can also induce exotic dynamical processes of dipoles, such as the combination of two neighboring dipoles to a new dipole with larger dipole moment. To be more specific,
 we first explore the dynamics of a "large dipole" that corresponds to the bound state of a particle and a next-nearest-neighbor hole whose dipole moment can be defined as $2$. Analogously, a "small dipole" denotes the bound state of a particle and a nearest-neighbor hole whose dipole moment can be defined as $1$. 
We show that a large dipole can be split into two small dipoles through its absorption of drive quanta from resonant tensor electric field.
The corresponding dipole dynamics is governed by two processes: splitting and recombination between a large dipole and two small dipoles, and tunneling of small dipoles, both of which depend on the amplitude of the driven potential. Furthermore, we extend this framework to the dynamics of dipoles with larger dipole moments. 
On the other hand, a dipole with a very large dipole moment is equivalent to two independent particle- and hole-type fractons, which means that our discussion is also relevant to individual fracton dynamics. We define the wave packet radius for dipoles and fractons to quantitatively characterize their expansion speed. This observable depends only on the local density distribution and can be measured directly in experiment using single atom resolution quantum gas microscopes ~\cite{bakr2009,sherson2010}.
To elucidate the dynamics of dipoles and an individual fracton, we map our model onto a lattice of hardcore bosons with density-dependent tunneling and nearest-neighbor interaction, the results are consistent with our system when the driven amplitude is not too large.

This work is organized as follows. In Sec. \ref{sec:model}, we introduce our model and give a brief discussion of time-dependent tensor electric fields. In Sec. \ref{sec:dipole dynamics}, we derive an effective model to analyze the dynamics of a large dipole with dipole moment $2$ under resonant driving, and demonstrate the agreement between the effective model and numerical simulations. In Sec. \ref{sec:fracton dynamics}, the discussion is generalized to broader scenarios to explain the dynamics of dipoles with larger dipole moments and an individual fraction in resonant tensor electric fields. In Sec. \ref{sec:Experimental setup}, an experimental realization of our scheme with ultracold atoms in optical lattices is proposed and the corresponding dynamics of dipoles and fractons  are predicted. Sec. \ref{sec:Conclusions} offers a discussion and outlook. 

\section{Model}
\label{sec:model}
We consider a one-dimensional dipole-conserving Bose-Hubbard model (DBHM) with a periodically driven quadratic potential, the Hamiltonian 
\begin{equation}\label{fullham}
    \hat{H}(t)=\hat{H}_0 + \hat{H}_e(t),
\end{equation}
can be written as 
\begin{eqnarray}\label{H0}
        &\hat{H}_0 = -J\sum_m (\hat{b}_{m-1}\hat{b}^{\dag 2}_m\hat{b}_{m+1} + h.c. ) + \frac{U}{2}\sum_m  \hat{n}_m(\hat{n}_m-1), \\
        &\hat{H}_e(t) = \frac{A}{2}\cos(\omega t) \sum_m m^2\hat{n}_m,
\end{eqnarray}
which correspond to the static and time-dependent components, respectively. In this Hamiltonian, $\hat{b}_m$ and $\hat{b}_m^\dag$ are bosonic annihilation and creation operators on site $m$, and $\hat{n}_m=\hat{b}^\dag_m\hat{b}_m$ is the local particle density operator, $J$ and $U$ are the strength of the correlated tunneling and repulsive on-site interaction, $\omega$ and $A$ correspond to the driving frequency and amplitude, respectively. 

Defining the dipole moment operator as $\hat{P}=\sum^L_{m=1}m\hat{n}_m$, the commutation relation $[\hat{P},\hat{H_0}]=0$ implies the conservation of the dipole moment in this system. Without the time-dependent part $\hat{H}_e(t)$, the Hamiltonian $\hat{H}_0$ only includes the correlated tunneling and on-site interaction. In the case of unit filling and $J\ll U$, the ground state of the system is a Mott insulator as $|\mathrm{MI}\rangle = \prod_m \hat{b}^\dag_m |0\rangle$. The fracton is defined as the excited state corresponding to an additional particle or hole on this ground state. For example, the state $|f_j\rangle = \hat{b}^\dag_j|\mathrm{MI}\rangle/\sqrt{2}$  with an additional particle at site $j$. It is easy to verify that the dynamics of this additional particle on site $j$ is constrained although it isn't the ground state of $\hat{H}_0$. The tunneling processes, led by correlated tunneling that conserves the dipole moment, will cost extra energy U, consequently the motion of the additional particle is frozen on the time scale of $\sim 1/J$ in the strongly interacting regime $J\ll U$.
It is worth pointing out that, in the regime of $J\sim U$, the fracton exhibits interesting dynamics~\cite{boesl2024}. 

\begin{figure}[htbp]
\includegraphics[width=1\columnwidth]{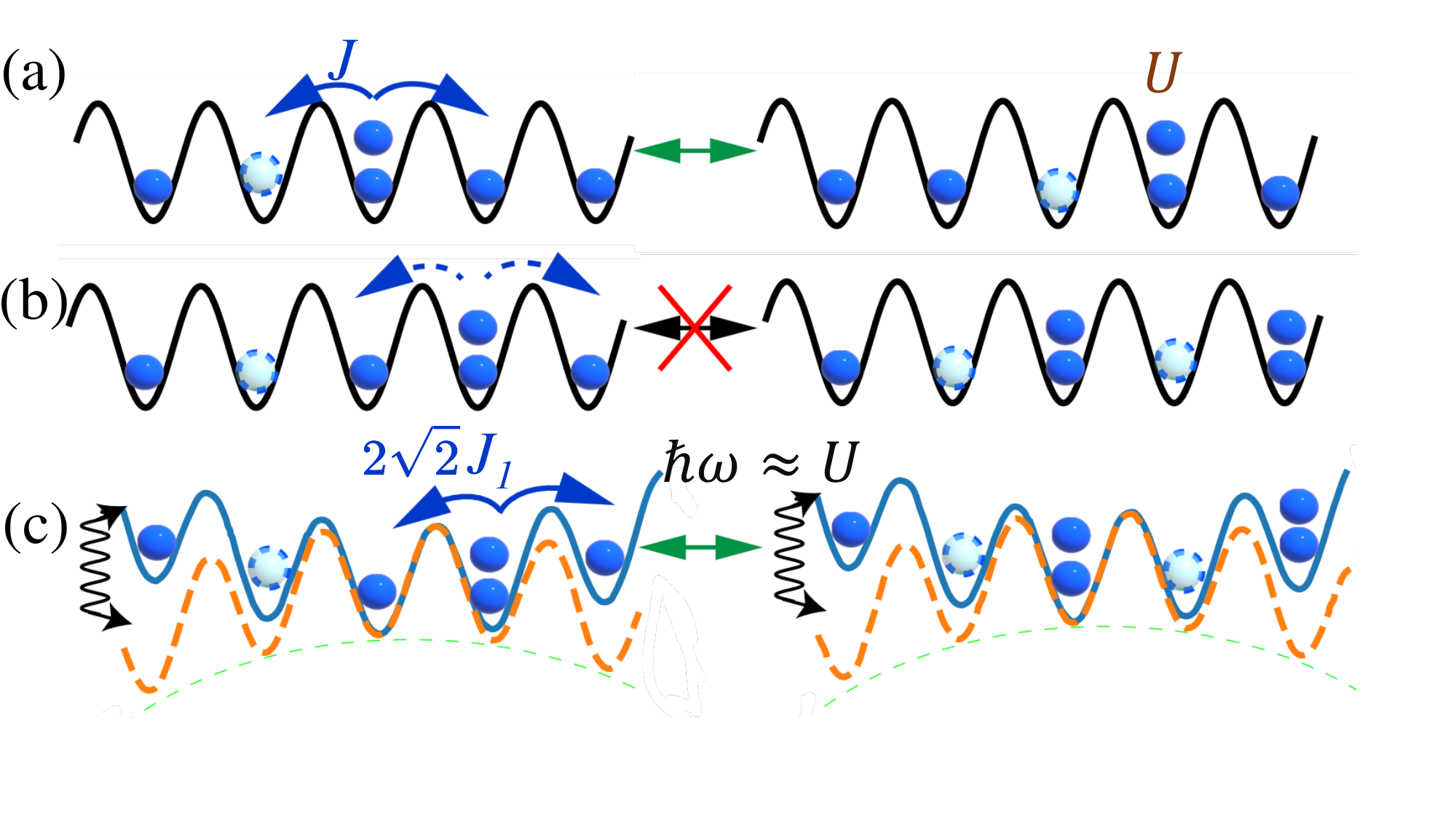}
\caption{Scheme of dipole dynamics in DBHM. (a) A small dipole can propagate freely through the lattice. (b) A large dipole with dipole moment $2$ is kinetically constrained because of the extra energy cost. (c) Schematic illustration of resonant driving: a time-dependent tensor electric field $E_{xx}$ introduced by periodically driven quadratic potential, resonantly mediates the splitting and recombination processes between dipoles with different moments, releasing large dipoles from their confinement.}
\label{fig:Scheme}
\end{figure}

On the other hand, the dipole defined in our work is a composite excitation combined by a particle and a hole-type fracton separated by $n$ lattice sites, as $\hat{D}^\dag_{j,n}|\mathrm{MI}\rangle =  \frac{1}{\sqrt{2}} \hat{b}_j\hat{b}^\dag_{j+n}|\mathrm{MI}\rangle$. Defining the dipole moment of an arbitrary state $|\psi\rangle$ as $P = \langle\psi|\hat{P}|\psi\rangle - \langle\mathrm{MI}|\hat{P}|\mathrm{MI}\rangle$, the dipole moment of $\hat{D}^\dag_{j,n}|\mathrm{MI}\rangle$ is $n$.
The dipole excitations in the system with Hamiltonian $\hat{H}_0$ can be classified into two different types by $n$. 
A small dipole, formed by a nearest-neighbor particle and hole excitation with dipole moment $n=1$, can move freely in the absence of a tensor electric field, and is expressed as $\hat{d}^\dag_j|\mathrm{MI}\rangle =  \frac{1}{\sqrt{2}} \hat{b}_j\hat{b}^\dag_{j+1}|\mathrm{MI}\rangle$. This mobility follows from the correlated-tunneling term in $\hat{H}_0$, which can be rewritten as $-2J\sum_m(\hat{d}^\dag_m\hat{d}_{m+1} + h.c.)$, thus revealing the dynamics of the small dipoles, as illustrated in Fig. \ref{fig:Scheme}(a). In contrast, a large dipole, corresponding to a particle and a hole separated by distance $n>1$, is prohibited to move on the time scale of $~1/J$ like single fractons, as shown in Fig. \ref{fig:Scheme}(b), and is defined as $\hat{D}^\dag_{j,n}|\mathrm{MI}\rangle =  \frac{1}{\sqrt{2}} \hat{b}_j\hat{b}^\dag_{j+n}|\mathrm{MI}\rangle$. Unlike the traditional dipole defined in textbook of electromagnetism, the dipoles here do not exhibit long-range interaction between the particle and hole, so as to the lack of long-range interaction between different dipoles. However, conservation of the dipole moment implies that the definition of dipole remains relevant.

These exotic physical phenomena imply that the DBHM is an important tool for exploring the dynamics of fractons and dipoles. More interesting thing are their dynamics under tensor gauge fields. As discussed in our earlier work, the system with Hamiltonian $\hat{H}_0$ is decoupled with vector gauge fields. Instead, a static tensor electric field can drive dipolar Bloch oscillation of small dipoles~\cite{zhang2025synthetic}. However, for an individual fracton and large dipoles, the static tensor electric field is still ineffective. As an example, the correlated tunneling can lead to splitting of a large dipole with dipole moment $2$ to two small dipoles, but the energy mismatch that arises from the on-site interaction suppresses this process, as shown in Fig. \ref{fig:Scheme}(b). Therefore, extra energy cost is required to drive a large dipole, and the periodically driven quadratic potential $\hat{H}_e(t)$, which corresponds to a time-dependent tensor electric field, is necessary. Because the kinetics of single particle is absent in Eq.~(\ref{H0}), the quadratic potential $\hat H_e(t)$ produces no dynamics for an isolated particle or hole. In contrast, it acts as an effective force on mobile, charge-neutral dipoles enabled by correlated tunneling, the strength of force is given by the second derivative of the potential, which is naturally interpreted as a rank-2 tensor gauge field.

Using the rotating frame $\hat{H}'(t) = \hat{\mathcal{U}}(t)\hat{H}(t)\hat{\mathcal{U}}^{-1}(t) + i\hbar(\partial_t\hat{\mathcal{U}}(t))\hat{\mathcal{U}}^{-1}(t)$, where the unitary operator $\hat{\mathcal{U}}$ is
\begin{equation}\label{unitaryops}
    \hat{\mathcal{U}}(t) = e^{i(\frac{A}{2\omega}\sin(\omega t) \sum_m m^2\hat{n}_m)/\hbar},
\end{equation}
the Hamiltonian $\hat{H}(t)$ can be translated into
\begin{eqnarray}\label{hamwithTEF}
\begin{aligned}
   \hat{H}'(t) =& -J\sum_m\left\{e^{-i\frac{A}{\hbar\omega}\sin(\omega t)} \hat{b}_{m-1}\hat{b}^{\dag 2}_m\hat{b}_{m+1} + h.c. \right\} \\
   &+ \frac{U}{2}\sum_m \hat{n}_m(\hat{n}_m-1). 
\end{aligned}
\end{eqnarray}
The additional phase $A_{xx}=-\frac{A}{\hbar\omega}\sin(\omega t)$ before the correlated tunneling term can couple to the second derivative of the bosonic field~\cite{zhang2025synthetic}. $\hat{H}'(t)$ is invariant under the transformation $\hat{b}^\dag_m\rightarrow\hat{b}^\dag_m e^{i\alpha_m}$ and $A_{xx}\rightarrow A_{xx}-2\alpha_m+\alpha_{m+1}+\alpha_{m-1}$. In the continuum limit, $A_{xx} \rightarrow A_{xx} + a^2\partial^2_x\alpha$, where $a$ is the lattice spacing. The second derivative in   this transformation identifies $A_{xx}$ as a rank-2 tensor gauge potential.

The complete tensor gauge field includes the spatial component $A_{xx}$ and the temporal component $A_0$. In Hamiltonian (\ref{fullham}), $A_{xx}=0$ and $A_0=\frac{A}{2a^2\hbar}\cos(\omega t) x^2$. The unitary transformation in Eq. (\ref{unitaryops}) can be viewed as a rank-2 tensor gauge transformation, $A_{xx}\rightarrow A_{xx}+ a^2\partial_x^2\alpha$ and $A_0\left(t\right)\rightarrow A_0\left(t\right)+\partial_t\alpha$, where $\alpha$ can be an arbitrary scalar field. So the physical consequence only depends on the gauge invariant $E_{xx}=a^2 \partial_x^2A_0-\partial_tA_{xx}$ \cite{You2021}, which corresponds to the tensor electric field. Choosing $\alpha=-\frac{A}{2a^2\hbar\omega}\sin(\omega t) x^2$ yields $A_0=0$ and $A_{xx}=-\frac{A}{\hbar\omega}\sin(\omega t)$, the rank-2 tensor electric field can be produced as
\begin{equation}\label{tensoracfield}
    E_{xx} = -\frac{\partial A_{xx}}{\partial t} = \frac{A}{\hbar}\cos(\omega t),
\end{equation}
where $A$ and $\omega$ are the strength and frequency of this tensor electric field, respectively. We should point out that, unlike the real gauge fields, this synthetic tensor gauge field don't satisfy Lorentz invariance, and the synthetic time-dependent electric field applied to neutral atoms will not lead to a corresponding magnetic field.

The engineering of the frequency $\omega$ leads to a series of intriguing phenomena~\cite{Weitenberg2021,sias2008observation}. For the extreme case where $\omega$ approaches zero, the time-dependent tensor electric field reduces to a static tensor electric field. The small dipoles will be driven by the electric field and the dipolar Bloch oscillation emerges. For the other extreme case where $\hbar\omega$ is much larger than any other energy scales and $A\ll \hbar\omega$, the effective time-independent Hamiltonian can be derived from high-frequency expansion ~\cite{eckardt2015high,goldman2014periodically,rahav2003effective,sias2008observation} as
\begin{eqnarray}\label{effectiveH}
\begin{aligned}
  H_\mathrm{hf} =& -J\mathcal{J}_0\left(\frac{A}{\hbar\omega}\right) \sum_m(\hat{b}_{m-1}\hat{b}^{\dag 2}_m\hat{b}_{m+1} + h.c. ) \\
  +& \frac{U}{2}\sum_m \hat{n}_m(\hat{n}_m-1),
\end{aligned}
\end{eqnarray}
where the strength of correlated tunneling is modulated by the $0$-th Bessel function $\mathcal{J}_0\left(\frac{A}{\hbar\omega}\right)$, which decreases as the driving amplitude $A$ increases. In both cases, the movement of large dipoles remains constrained. 

Therefore, we focus on the resonant dynamics where $U\approx \hbar\omega$, in which the tensor electric field leads to splitting and recombination of dipoles, thereby enabling mobility of the large dipole, as shown in Fig. \ref{fig:Scheme}(c). This mechanism can be analogized to the role of time-dependent electromagnetic fields in optical Feshbach resonance, where molecules can dissociate into their constituent atoms by absorbing or emitting one photon~\cite{chin2010feshbach,Theis}. 

\section{Dipole dynamics under resonant driving}
\label{sec:dipole dynamics}
We investigate dipole dynamics in the strong-coupling regime ($U\gg J$) at unit filling, where the driving frequency is near-resonant with the on-site interaction as $\hbar\omega\approx U$. For a general case, the time-dependent coefficient in front of the correlated tunneling term in Eq. (\ref{hamwithTEF}) can be expanded as $\exp\big\{i\frac{A}{\hbar\omega}\sin(\omega t)\big\} = \sum^\infty_{k=-\infty} \mathcal{J}_k
\left(\frac{A}{\hbar\omega}\right) e^{i k\omega t}$, corresponding to the $k$ photon-assisted correlated tunneling, with the $k$-th order Bessel functions $\mathcal{J}_k\left(\frac{A}{\hbar\omega}\right)$ as amplitudes. In the near-resonant case, these photon-assisted correlated tunneling will play a crucial role in the dynamical processes. 

In this section, we will explore the dynamics of a large dipole with dipole moment $2$ as an example. We omit $n$ in the dipole creation operator $\hat{D}_{j,n}$ as $\hat{D}_j$ when $n=2$ for simplicity. The initial state is prepared as $|\psi_0\rangle = \hat{D}^\dag_j|\mathrm{MI}\rangle$ with a large dipole at the $j$ site.       
Analytical calculation of the dynamics of this many-body system is not easy. However, Floquet analysis of a four-site lattice can provide an intuitive picture of the dynamics. The details of the analysis are presented in the Appendix \ref{sec:four-site}, we just show the main results here. The tunneling of a large dipole is restricted on the time scale of $\sim 1/J$ because the corresponding tunneling amplitude can be estimated as $\sim J^2/U$, which is much smaller than $J$. But a large dipole can be split into two small dipoles and recombine via one-photon assisted correlated tunneling with amplitude $J_1=J\mathcal{J}_1\left(\frac{A}{\hbar\omega}\right)$. When the driven frequency is resonant with on-site interaction, $\hbar\omega \approx U$, this splitting and recombination process is drastically enhanced and plays a crucial role in the dipole dynamics.

For a sufficiently large lattice,the splitting and recombination processes are still maintained. Unlike the four-site lattice, these two small dipoles can move freely with the amplitude as $J_0=J\mathcal{J}_0\left(\frac{A}{\hbar\omega}\right)$, except for the exchange of each other. Furthermore, multiphoton-assisted correlated tunneling processes related to the $k$-th Bessel function with $k>1$ also emerge, with the amplitude as $J_k=J\mathcal{J}_k\left(\frac{A}{\hbar\omega}\right)$. However, when the driven amplitude $A$ is not so large, these terms can be ignored, then the dipole dynamics include not only the splitting of large dipoles, but also the nearest-neighbor tunneling of small dipoles,
whose amplitudes are determined by $J_1$ and $J_0$, respectively. 

To build intuition for dipole dynamics and provide an independent cross-check of the original Hamiltonian (\ref{fullham}), we construct an effective model. However, the original Hamiltonian (\ref{fullham}) remains our primary object of study and is valid throughout. Defining the Fock states $|{\bf 1}_m {\bf 1}_{m^\prime}\rangle = \hat{d}^\dag_m \hat{d}^\dag_{m^\prime }|\mathrm{MI}\rangle$ as two small dipoles on sites $m$ and $m^\prime$ where $m^\prime-m \ge 2$, and $|{\bf 1}_m {\bf 1}_{m+1}\rangle = \hat{D}^\dag_m |\mathrm{MI}\rangle$ as one large dipole with dipole moment $2$ on site $m$, the dipole dynamics can be described using the effective Hamiltonian $\hat{H}_D = \hat{H}_{D,\mathrm{bulk}} + \hat{H}_{D,\mathrm{edge}}$, where the bulk term is

\begin{eqnarray}\label{effhamdipole}
\begin{aligned}
   \hat{H}_{D,\mathrm{bulk}} &=  -\frac{4J^2_0}{U}\sum_m |{\bf 1}_m {\bf 1}_{m+1}\rangle \langle {\bf 1}_{m+1} {\bf 1}_{m+2}| \\
   &- 2\sqrt{2}J_1\sum_m |{\bf 1}_m {\bf 1}_{m+1}\rangle \langle {\bf 1}_m {\bf 1}_{m+2}|\\ &+\sqrt{2}J_1\sum_m |{\bf 1}_m {\bf 1}_{m+1}\rangle \langle {\bf 1}_{m-1} {\bf 1}_{m+1}| \\
   &+ \delta^\prime\sum_m |{\bf 1}_m {\bf 1}_{m+1}\rangle \langle {\bf 1}_m {\bf 1}_{m+1}|\\
   -2J_0& {\sum_{m,m^\prime}}^\prime |{\bf 1}_m {\bf 1}_{m^\prime}\rangle (\langle {\bf 1}_{m+1} {\bf 1}_{m^\prime}| + \langle {\bf 1}_m {\bf 1}_{m^\prime+1}|) + h.c,
\end{aligned}
\end{eqnarray}
where $\delta^\prime = \delta-\frac{20J^2_0}{U}\approx \delta$ is the effective energy detuning of the large dipole, the energy correction $-\frac{20J^2_0}{U}$ comes from the second-order perturbation. (More details are discussed in Appendix \ref{sec:more-site}.) The prime in the summation indicates that tunneling terms for $m^\prime-m\le 2$ are excluded. The edge term $\hat{H}_{D,\mathrm{edge}} = \frac{2J^2_0}{U} |{\bf 1}_1 {\bf 1}_2\rangle \langle {\bf 1}_1 {\bf 1}_2| + \frac{8J^2_0}{U} |{\bf 1}_{L-2} {\bf 1}_{L-1}\rangle \langle {\bf 1}_{L-2} {\bf 1}_{L-1}|$ is irrelevant when the length of the lattice is large enough.

Exploiting the fact that two dipoles on the same site are energetically suppressed, we map dipoles to hardcore bosons. The effective Hamiltonian (\ref{effhamdipole}) thus describes two hardcore bosons on a lattice,
where the state $|{\bf 1}_m {\bf 1}_m^\prime\rangle$ describes a configuration with one boson occupying site $m$ and the other occupying site $m^\prime$. 
Consequently, the dynamical evolution of the system can be effectively treated as the expansion dynamics of two hardcore bosons with the initial state as $|{\bf 1}_j {\bf 1}_{j+1}\rangle$. In this effective model, the splitting of large dipole and the movement of two small dipoles has been mapped onto density-dependent tunneling, which means that the tunneling amplitude of a hardcore boson depends on the occupation of bosons on its neighbor site. $\delta'$ can be considered as the strength of nearest-neighbor interaction. Thus, this effective Hamiltonian (\ref{effhamdipole}) for hardcore bosons represents a complex dynamic process which is difficult to be solved analytically. 

To prove the validity of the discussion, we employ two different methods to simulate this expansion process independently: time-evolving-block-decimation (TEBD) method based on the original Hamiltonian (\ref{fullham}) using the TenPy library~\cite{hauschild2018efficient}, and exact diagonalization of the effective Hamiltonian (\ref{effhamdipole}). To maintain consistency between these two methods, in our TEBD simulations we impose a local Hilbert space truncation at the maximum occupation number $N_\mathrm{max}=2$ to suppress multiphoton-assisted tunneling processes~\cite{Zhang2025CompAppendix}. The bond dimension 
$\chi_{max}$ is 80, and time step $\delta t=0.001\hbar/J$. The simulation results show that the outcomes of both methods are in good agreement, therefore, we just show the results of TEBD based on the original Hamiltonian (\ref{fullham}). Throughout, unless otherwise specified, all numerical results reported here are from TEBD.

\begin{figure*}[htbp]
\includegraphics[width=1.8\columnwidth]{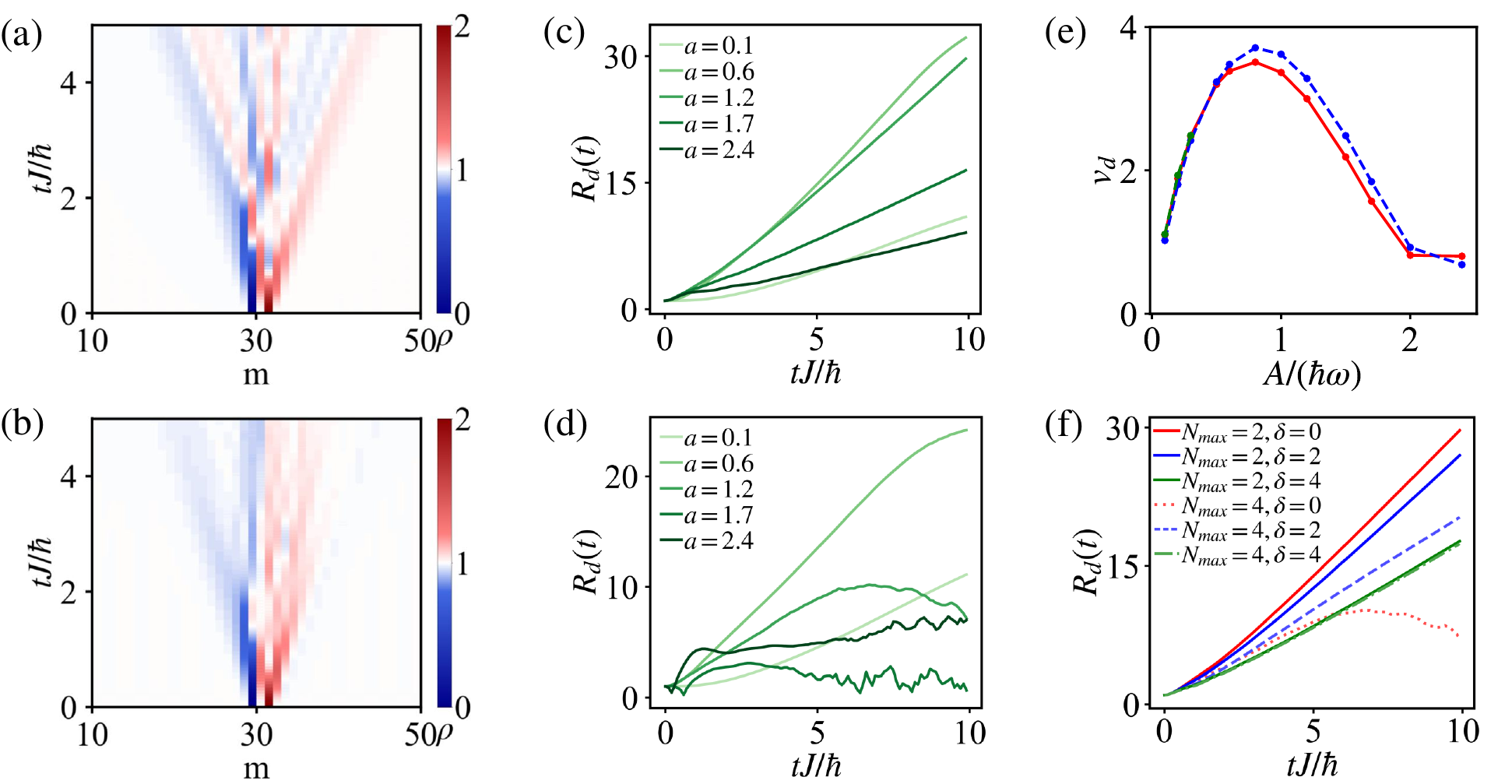}
\caption{Dynamics of resonantly driven dipole excitations in the one-dimensional lattice of $L=61$. (a), (b) Time evolution of particle density $\rho_j(t)$ obtained with TEBD at resonance $\hbar\omega = U = 120 J$, and driving strength $A/(\hbar\omega)=1.2$, with maximum occupation number (a) $N_{max}=2$, (b) $N_{max}=4$. (c), (d) The evolution of wave packet radius $R_d(t)$ for various driving amplitudes $A$, parameterized by $a=A/(\hbar\omega)$, with (c) $N_{max}=2$, (d) $N_{max}=4$. (e) Expansion speed $v_d=\frac{dR_d(t)}{dt}$ for various driving amplitudes $A$. The Red solid line and green dashed line are obtained from the TEBD data in (c) and (d) respectively, while the blue dashed line is obtained from the result of effective Hamiltonian (\ref{effhamdipole}). The points corresponding to large $A$ of the green curve are excluded when $R_d(t)$ becomes non-ballistic. (f) The evolution of wave packet radius $R_d(t)$ at $A/(\hbar\omega)=1.2$ for detunings $\delta=\hbar\omega-U$, with $\delta/J=0$ (red), $2$ (blue) and $4$ (green). Solid lines correspond to maximum occupation number $N_{max}=2$; Dashed lines to $N_{max}=4$.}
\label{fig:L60 evolution}
\end{figure*}
 
In the resonant case, the local charge density $\rho^d_m = \langle \hat{n}_m \rangle -1$ undergoes an approximately ballistic expansion with the speed depending on the driving amplitudes. Fig. \ref{fig:L60 evolution}(a) shows the expansion at $A=1.2\hbar\omega$ as an example. As the local dipole moment can be defined as $p_m=(m-L/2)\rho^d_m$, the expansion speed of dipoles can be characterized by the wave packet radius of dipoles, which can be defined as
\begin{equation}\label{wavepacketradius2}
  R^2_d(t) = \sum_m (m-L/2)^2 p_m(t)/P = \sum_m (m-L/2)^3 \rho^d_m(t)/P,
\end{equation}
where $P = \langle\psi|\hat{P}|\psi\rangle - \langle\mathrm{MI}|\hat{P}|\mathrm{MI}\rangle$ is the dipole moment. Fig. \ref{fig:L60 evolution}(c) shows the evolution of the wave packet radius $R_d(t)$ for different values of the driving amplitude $A$. 
The linear increase of $R_d(t)$ verifies the ballistic expansion of the dipoles. The expansion speed $v_d$ is estimated by a linear fit to $R_d(t)$ versus time, and its dependence on $A$ is shown as the solid red line in Fig. \ref{fig:L60 evolution}(e). The result of the exact diagonalization using the effective Hamiltonian (\ref{effhamdipole}) is shown as the dashed blue line in Fig. \ref{fig:L60 evolution}(e), which is consistent with our TEBD simulation. The expansion speed is governed by two distinct processes simultaneously: the splitting of a large dipole and the tunneling of small dipoles. When $A$ is very small, although the tunneling strength $J_0$ for small dipoles is large, the initial state is prepared as a large dipole whose splitting is determined by the small $J_1$, so the expansion speed is primarily controlled by $J_1$. With an increase of $A$, the values of $J_0$ and $J_1$ are comparable and tunneling of the small dipole starts to be important. As $A$ increases to $A/(\hbar\omega) \approx 2.4$, the zero point of $J_0$, the evolution is dominated by the splitting of the large dipole, and the expansion speed depends only on $J_1$. 
 
In the above TEBD simulations, we neglect multiphoton-assisted correlated tunneling terms for simplicity. However, these terms should be taken into account when the driving amplitude $A$ is large enough that higher-order Bessel functions $\mathcal{J}_k\left(\frac{A}{\hbar\omega}\right)$ are no longer negligible. To incorporate higher-order photon-assisted correlated tunnelings, we increase the maximum local occupation number to $N_\mathrm{max}=4$. As shown in Fig. \ref{fig:L60 evolution} (b), when the driving amplitude is $A=1.2\hbar\omega$, the evolution of local charges deviates from the previous results with $N_\mathrm{max}=2$. On the time scale of $\sim 1/J$, the ballistic expansion persists with a deviation of the expansion speed from the previous results. However, if $A$ go on increasing, the evolution of $R_d(t)$ is no longer linear, indicating that multiphoton-assisted correlated tunnelings exert notable influence, as shown in Fig. \ref{fig:L60 evolution} (d). The green dash line in Fig. \ref{fig:L60 evolution} (e) shows the speed of ballistic expansion as a function of $A$, which is only relevant when $A$ is small. When $A$ is too large, this speed is meaningless because the evolution of $R_d(t)$ is no longer linear anymore.

To mitigate the effects of multiphoton-assisted correlated tunneling, we can employ a near-resonant condition with finite $\delta$ instead of exact resonance. In this case, the energy detuning for the $n$-th photon-assisted correlated tunneling term is magnified by $n\delta$. As shown in Fig. \ref{fig:L60 evolution} (f), the ballistic expansion is restored, and the results are consistent with the simulation of the effective Hamiltonian in a larger range of $A$. 

\section{More general dipole dynamics and fracton dynamics}
\label{sec:fracton dynamics}
The above discussion on the dynamics of a dipole with dipole moment of $2$ can be extended to more general cases. For a large dipole with dipole moment larger than $2$, the dynamical processes also include its splitting into a series of small dipoles and tunneling of these small dipoles. On the other hand, a large dipole with a very large dipole moment can be regarded as a pair of independent particle- and hole-type fractons separated by a long distance. Consequently, the dynamics of a large dipole can be viewed as the individual movement of these fractons and their interference. It implies that the time-dependent tensor electric field can break the mobility constraint of an independent particle- or hole-type fracton through photon-assisted correlated tunneling, and thereby induces interesting fracton dynamics. 

We first investigate the dynamics of a dipole excitation $\hat{D}^\dag_{j,n}|\mathrm{MI}\rangle = \frac{1}{\sqrt{2}}\hat{b}_j\hat{b}^\dag_{j+n}|\mathrm{MI}\rangle$ on a unit-filled Mott insulator state $|\mathrm{MI}\rangle = |11\cdots 1\rangle$, where $n\ge 3$.
Without the time-dependent tensor electric field, the tunneling of this large dipole can only be achieved by a $q$-th order process with the amplitude $J_{q,D}\sim J^q/U^{q-1}$, which can be ignored on a time scale of $1/J$. Within the framework discussed above and excluding the multiphoton-assisted correlated tunneling, the dynamics of this large dipole arises only from its splitting into two or more smaller dipoles. For example, a large dipole with dipole moment $n$ at site $j$ can split into two adjacent dipoles with dipole moment $n-1$ and $1$ with photon-assisted correlated tunneling, while two final states can be expressed as $\hat{D}^\dag_{j,n-1}\hat{d}^\dag_{j+n}|\mathrm{MI}\rangle$ and $\hat{d}^\dag_{j-1}\hat{D}^\dag_{j+1,n-1}|\mathrm{MI}\rangle$. 
Then the cascade of the splitting process eventually dissociates a large dipole into a series of small dipoles, where the summation of their dipole moments is conserved. 

Although the total dipole moment is conserved, the splitting processes fall into two distinct categories. The first changes the dipole count by $1$, for example, a large dipole with moment $n$ split into two adjacent dipoles with dipole moments $n-1$ and $1$.
These processes come from photon-assisted correlated tunneling, and the amplitudes are proportional to $J_1$. The other preserves the dipole count, for instance, from two adjacent dipoles with moments $n-1$ and $1$ to two dipoles with moments $n-2$ and $2$. These processes arise from static correlated tunneling whose amplitudes are proportional to $J_0$. Following the preceding discussion based on the effective Hamiltonian Eq. (\ref{effhamdipole}), this system can be mapped onto an equivalent hardcore boson lattice model comprising $n$ bosons with density-dependent tunneling and nearest-neighbor interaction. The $n$-boson system does not differ qualitatively from the two-boson scenario, and the detail is shown in the Appendix \ref{sec:appendix}. Our analysis shows that the resulting dipole wave packet exhibits nearly the same expansion behavior as the dipole with dipole moment $2$. 
TEBD simulations indicate that the expansion velocity of these large dipoles is comparable to that of the dipole with moment $2$.

The more interesting thing is the dynamics of an individual fracton. 
A representative example is the preparation of an independent particle-type fracton excitation $|f_j\rangle = \hat{b}^\dag_j |\mathrm{MI}\rangle$ as the initial state. Consequently, the filling factor $\langle \rho \rangle \ne 1$. Driven by a near-resonant time-dependent tensor electric field, the fracton can move by absorbing the energy from the field, which accompanied by creating an additional dipole, thereby preserving the total dipole moment. As in the above discussion, a similar effective lattice model of hardcore bosons can be constructed and the details can be found in the Appendix \ref{sec:appendix}.

\begin{figure}[htbp]
\includegraphics[width=1\columnwidth]{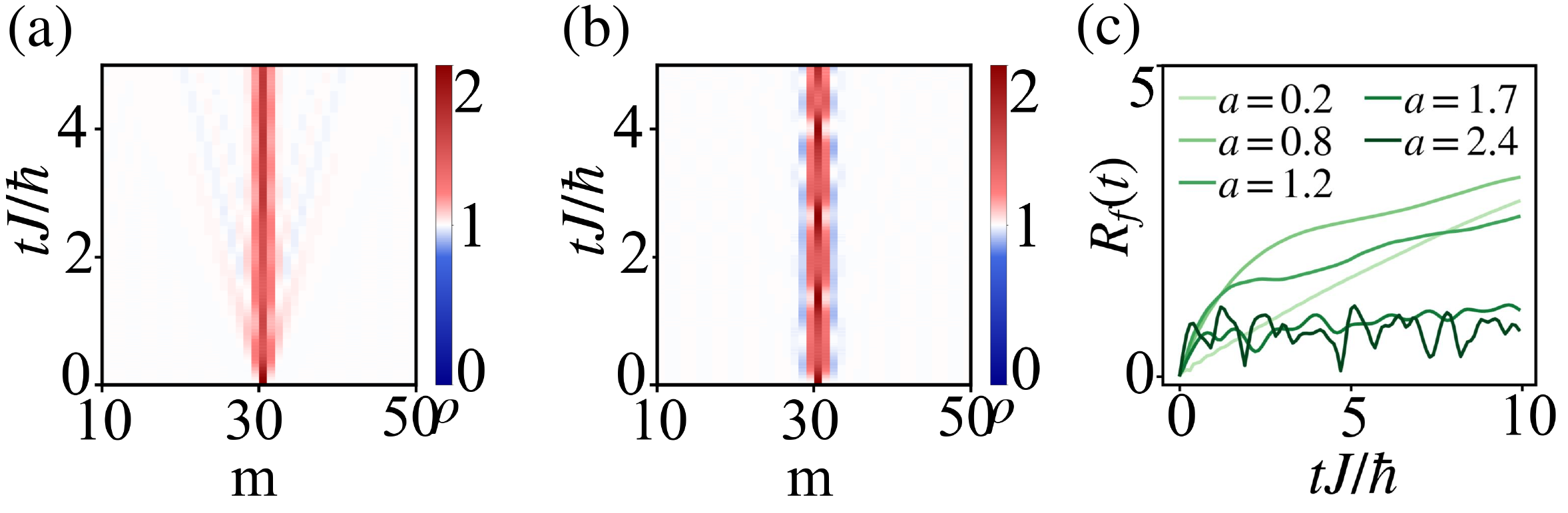}
\caption{Dynamics of a near-resonantly driven individual fracton excitation in the one-dimensional lattice of $L=61$. (a), (b) Time evolution of particle density $\rho_j(t)$  at detuning $\delta=\hbar\omega - U = 4 J$ with $U=120J$, and (a) $A/(\hbar\omega)=1.2$, (b) $A/(\hbar\omega)=2.4$. (c) The evolution of wave packet radius $R_f(t)$ for various driving amplitudes $A$, parameterized by $a=A/(\hbar\omega)$.}
\label{fig:fracton}
\end{figure}

The results of TEBD simulation are shown in Fig. \ref{fig:fracton}. To suppress multiphoton-assisted correlated tunneling, we introduce a finite energy detuning, which slows down the expansion speed of the fracton, as shown in Fig. \ref{fig:fracton}(a). Unlike dipole excitations, the wave packet radius of this individual fracton needs to be characterized by
\begin{equation}\label{wavepacketradius3}
    R^2_f(t) = \sum_m (m-L/2)^2 \rho^d_m(t).
\end{equation}
The evolution of the wave packet radius $R_f(t)$ for various driving amplitudes $A$ is shown in Fig. \ref{fig:fracton}(c). The ballistic expansion can still be verified by the linear increase of $R_f(t)$ except at very early times. A special case occurs for sufficiently large $A$ (near $A/(\hbar\omega) = 2.4$), where the density oscillates rather than expands, as shown in Fig. \ref{fig:fracton}(b). In this regime, all the processes proportional to $J_0$ are frozen, not only the tunneling of small dipole, but also the reorganization of large dipoles which keeps the number of dipoles. It prevent the cascade of the splitting and only the oscillation can be observed. 

\section{Experimental realization}
\label{sec:Experimental setup}
In experiment, a Bose-Hubbard model with a strong linear tilt described by
\begin{eqnarray}\label{tiltedBHM}
\begin{aligned}
 \hat{H}_\mathrm{tilted} =& -J^{\prime}\sum_m(\hat{b}^\dag_m\hat{b}_{m+1} + h.c.) \\
 +& \frac{U}{2}\sum_m\hat{n}_m(\hat{n}_m-1) + \Delta\sum_m m\hat{n}_m,
 \end{aligned}
\end{eqnarray}
can be considered as an accessible platform to implement dipole-conserving dynamics. Here, $\Delta$ is the energy detuning between nearest neighbor sites. In the limit $J^{\prime}, U\ll \Delta$, the single-particle tunneling is prohibited, and $\hat{H}_0$ in the original Hamiltonian (\ref{fullham}) can be derived from (\ref{tiltedBHM}) using the perturbation approach, where the amplitude of the correlated tunneling can be estimated as $J = -J^{\prime 2}\big(\frac{1}{U+\Delta} - \frac{1}{\Delta}\big)\approx J^{\prime 2}U/\Delta^2$. If the parameters of Hamiltonian (\ref{tiltedBHM}) is chosen as $J^{\prime} \ll \Delta$, the ratio $J^{\prime}/U \approx (J^{\prime}/\Delta)^2$ is a second-order small quantity, which is relevant with our above discussion. The model and the
required parameter regime are readily accessible in optical
lattice experiments~\cite{gross2017}. The linear tilt can be constructed by applying a magnetic field with uniform gradient~\cite{Bloch2013}. The value of $\Delta/h$ in ultracold atomic experiment could be tuned up to 10kHz, much larger than $J^{\prime}/h$ and $U/h$, which are typically of the order of a few tens Hz and a few hundreds Hz~\cite{Bloch2008}. So our model in Eq.(\ref{tiltedBHM}) is thus directly accessible in current experiments.

\begin{figure}[htbp]
\includegraphics[width=1\columnwidth]{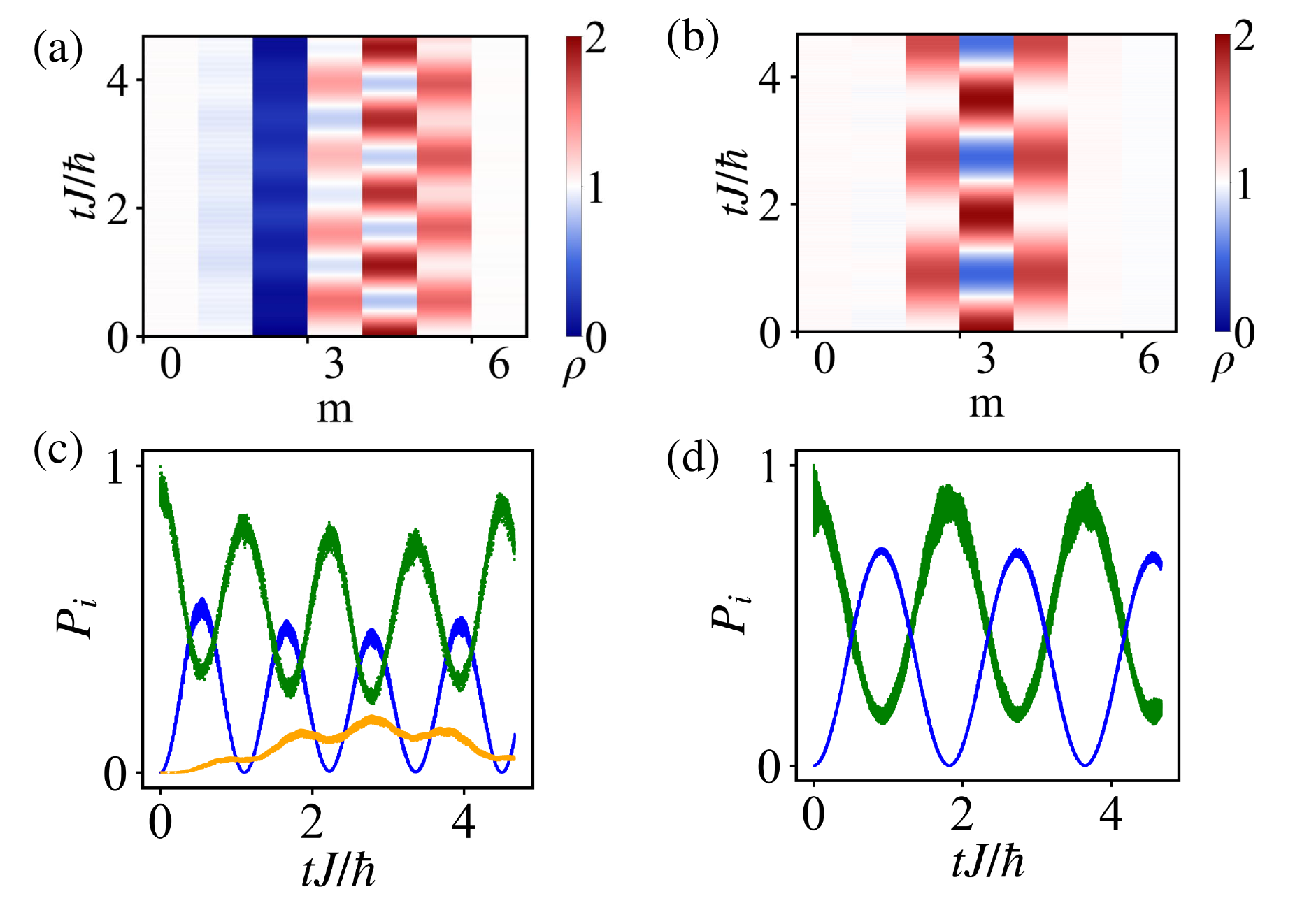}
\caption{Numerical simulation for the system simultaneously subjected to a strong linear tilt potential and a periodically driven quadratic potential whose driving frequency $\omega$ is tuned close to resonance with the on-site interaction $U$. Time evolution of particle density $\rho_j(t)$ is shown for $U=\sqrt{5}J^{\prime}$ and $\Delta=12\sqrt{3}J^{\prime}$. (a) Time evolution of a dipole excitation in one-dimensional lattice of $L=7$, with $\hbar\omega = 0.8989U$ and $A/(\hbar\omega)=1.2$. The minimum of the quadratic confinement is centered at site  $(L-3)/2$. (b) Time evolution of a fracton excitation in the one-dimensional lattice of $L=7$, with $\hbar\omega = 0.9677U$ and $A/(\hbar\omega)=1.2$. The quadratic potential is centered at site $(L-1)/2$.  (c) Time evolution of the state populations $P_{i}=|\langle \psi(t)|\phi_{i}\rangle|^2$ for the basis states $|\phi_1\rangle=|1101211\rangle$, $|\phi_2\rangle=|1102021\rangle e^{-i\omega t}$, $|\phi_3\rangle=|1110121\rangle$. Green, blue, and orange curves correspond to $P_1$, $P_2$ and $P_3$, respectively. (d) Time evolution of the state populations $P_{i}=|\langle \psi(t)|\phi_{i}\rangle|^2$ for the basis states $|\phi_1\rangle=|1112111\rangle$, $|\phi_2\rangle=|1120211\rangle$. Green and blue curves correspond to $P_1$ and $P_2$, respectively.}
\label{fig:tilted BHM}
\end{figure}

In this work, we rely on the simultaneous presence of the correlated tunneling term and the periodically driven quadratic potential in Eq.~(\ref{fullham}). The time-dependent potential can be engineered by exploiting a perturbative periodically modulated harmonic potential. Such modulation of harmonic traps has been experimentally demonstrated in ultracold atomic systems by amplitude-modulating laser beams~\cite{Jiang2023}.
However, the coexistence of a linear tilt and a time-dependent quadratic driving makes the many-body dynamics more intricate. With the unitary operator $\hat{\mathcal{U}}'(t)=\exp \left[-i\left(\Delta t \sum_m m \hat{n}_{m}+\frac{A}{2 \omega} \sin (\omega t) \sum_m m^{2} \hat{n}_{m}\right)/\hbar\right]$, the Hamiltonian $\hat{H} = \hat{H}_\mathrm{tilted}+\hat{H}_e(t)$ can be rewritten in rotating frame as
\begin{eqnarray}\label{tiltedBHMrot}
\begin{aligned}
    \hat{H}'(t) &= -J^{\prime} \sum_{m}\left[e^{i\left(\Delta t+\frac{A}{\omega} \sin (\omega t)\left(m+\frac{1}{2}\right)\right)/\hbar} \hat{b}_{m}^{\dagger} \hat{b}_{m+1}+\text { h.c. }\right]\\&+\frac{U}{2} \sum_{m} \hat{n}_{m}\left(\hat{n}_{m}-1\right),
 \end{aligned}
\end{eqnarray}
where the tunneling phase is modulated by two incommensurate driving frequencies,  $\Delta$ and $\omega$, the Hamiltonian becomes quasi-periodic in time, so the usual high-frequency expansion method for large $\Delta$ breaks down. 
A qualitative description with multi-mode Floquet theory~\cite{Martin,Nathan} shows that the effective Hamiltonian contains many other terms, such as the nearest-neighbor  interaction of small dipoles and spatial-dependent energy detuning, whose magnitude even larger than the correlated tunneling terms, thereby breaking down the description of our effective model. 
Nevertheless, the simulation with TEBD can still provide a numerical tool for exploring the dipole and fracton dynamics in this complex system. Numerical simulations show that, when the driving frequency $\omega$ is tuned to be near-resonant with the on-site interaction ($\hbar\omega \approx U$), the splitting and recombination of the large dipoles with moment $2$ can still be observed, as shown in Fig. \ref{fig:tilted BHM}(a). But instead of the system with only correlated tunneling, the expansion of two neighbor small dipoles is prevented, and the system undergoes an oscillation between a large dipole and two nearest-neighbor small dipoles on a time scale of $\sim 1/J$, as shown in Fig. \ref{fig:tilted BHM}(c). Furthermore, the oscillation can also be observed in fracton dynamics by fine-tuning the driving frequency $\omega$, as in Fig. \ref{fig:tilted BHM}(b) and (d). 

All of these simulations show that the periodically driven quadratic potential remains available to modulate the dipole and fracton dynamics. However, achieving more sophisticated dynamical control will require the design and implementation of more intricate modulation sequences. 
Recently, dipole dynamics in a strongly tilted optical lattice has been observed experimentally \cite{kim2025Multi}. Using the parameters in experiment, the correlated tunneling amplitude $J/h$ can be estimated as about $1.7\mathrm{Hz}$, which means that the time scale of dipole or fracton dynamics is about $\hbar/J\approx 100\mathrm{ms}$. Comparing with the prethermal timescale in \cite{kim2025Multi}, our numerical simulation shows the dipole and fracton dynamics can be visible during this period. This indicates that coherent evolution in our scheme is experimentally feasible.

Finally, we should point out that Bose-Hubbard model with a strongly linear tilted potential in Eq. (\ref{tiltedBHM}) provides a unified platform for dipole-conserved dynamics \cite{kim2025Multi} and 
a series of other important quantum phenomena, such as the Stark many-body localization (SMBL) \cite{Berislav2022,Berislav2024}.
Unlike conventional disorder-induced many-body localization (MBL), SMBL arises in a strongly tilted interacting system, energy conservation effectively suppresses individual particle motion and leads to the conservation of the dipole moment \cite{kim2025Multi}.
SMBL shares many similarities with MBL, such as the logarithmical grows of entanglement entropy after a quench \cite{Schulz2019,Taylor2020,Berislav2022,Berislav2024}. The dipole-conserving dynamics discussed here and SMBL originate from the same mechanism in a strongly tilted interacting system,
which implies profound intrinsic connections between these two topics. For example, both dipole conservation and repulsive interaction play crucial roles. 

Even so, there are still some distinctions between these two topics. First, our work focuses on the short-time coherent dynamics of initial states with only one or a few dipoles, 
whereas SMBL is characterized by slow dynamics at long times, most notably a logarithmic growth of entanglement entropy persisting over extended time evolution.
 Second, the quadratic potential plays the role of synthetic tensor electric field, which is indispensable for coherent controlling the dipole dynamics. In SMBL, an additional quadratic potential is only used to lift the degeneracies of 
 states with the same center of mass \cite{Schulz2019}, which don't change the intrinsic physics qualitatively. 

The periodically driving potential for a quantum many-body system may lead to destabilizing or heating for a long-time evolution, but numerical simulation shows that the quantum coherence is maintained in the time scale we considered, as in Fig. \ref{fig:tilted BHM}. 
 It is instructive to further investigate the deeper connection between SMBL and dipole-conserved dynamics in the future.
 
\section{Conclusions and outlook}
\label{sec:Conclusions}
Using a periodically driven quadratic potential, we introduce a time-dependent tensor electric field in the one-dimensional DBHM. 
Under near-resonant driving, the photon-assisted correlated tunneling not only leads to the splitting of a large dipole but also drives an individual fracton. Furthermore, the expansion of dipoles and fractons in this resonant time-dependent tensor electric field has been analyzed.
We find that on the time scale of the inverse of correlated tunneling amplitude, the dipole and fracton dynamics exhibit roughly ballistic expansion, and the expansion speed is primarily governed by the splitting of large dipoles and the tunneling of small dipoles, both of whose amplitudes depend on the tensor electric field. We also mapped these dynamical processes to the dynamics of a few hardcore bosons in a one-dimensional lattice with density-dependent tunneling, which provides an equivalent comprehension for the dipole and fracton dynamics.

Our work establishes a theoretical framework for simulating the time-dependent tensor gauge fields and their coupling to fracton phases of matter. 
In the future, we will extend the Floquet engineering scheme based on the tensor gauge field to higher dimensions, thereby enabling the manipulation of fracton excitations such as lineons and planons. Furthermore, we will explore non-equilibrium dynamics driven by time-dependent tensor gauge fields, with a focus on phenomena like subdiffusive transport and the emergence of dynamical scar states in driven systems. These efforts promise to deepen our understanding of quantum many-body physics.

{\bf Acknowledgements}: This work is supported by National Natural Science Foundation of China (Grant No.12174138),  and the Quantum Science and Technology - National Science and Technology Major Project (Grant
 No. 2021ZD0302300). The TEBD simulations were performed by using the Tensor Network Python (TeNPy) package developed by J. Hauschild and F. Pollmann~\cite{hauschild2018efficient} and were run on the HPC Platform of Huazhong University of Science and Technology.

\begin{appendix}
\section{Resonant dynamics in a four-site lattice}
\label{sec:four-site}
The employment of Floquet theory is necessary for exploring the detail of dynamic processes. We will first provide a brief review of the Floquet theory~\cite{eckardt2015high}. Consider the Schr\"{o}dinger equation $i\hbar\partial_t|\psi(t)\rangle = \hat{H}(t)|\psi(t)\rangle$ where the time-dependent Hamiltonian $\hat{H}(t) = \hat{H}(t+T)$ is periodic with period $T=2\pi/\omega$. Floquet theory states that the solutions should take the form $|\psi_n(t)\rangle = |u_n(t)\rangle e^{-i\epsilon_n t/\hbar}$ with periodic Floquet modes $|u_n(t)\rangle = |u_n(t+T)\rangle$ and $\epsilon_n$ is the corresponding quasi-energy. The time-periodicity implies that quasi-energies are defined modulo $\hbar\omega$, such that they can be rewritten as $\epsilon_{mn} = \epsilon_n +m\hbar\omega$, where $m$ is an integer, and the corresponding Floquet modes are given by $|u_{mn}(t)\rangle = |u_n(t)\rangle e^{im\omega t}$. They satisfy the Schr\"{o}dinger equation as
\begin{equation}\label{floquettheory}
    (\hat{H}(t) - i\hbar\partial_t)|u_{mn}(t)\rangle = \epsilon_{mn} |u_{mn}(t)\rangle,
\end{equation}
which provides a framework for understanding the dynamics under periodic driving. By choosing a complete set of orthonormal basis states $|\alpha m \rangle = |\alpha\rangle e^{im\omega t}$ in the extended Hilbert space,  where $\{|\alpha\rangle\}$ forms the basis of the Hilbert space of the original Hamiltonian (\ref{H0}), the dynamic process can be analyzed using the Floquet matrix $M$, which is defined as $M_{\alpha m,\beta n} = \langle\alpha m|(\hat{H}(t) - i\hbar\partial_t)|\beta n\rangle$. By diagonalizing this Floquet matrix with a finite size cut-off, the quasi-energies and the corresponding Floquet modes can be estimated.

In the lattice with only four sites, the Fock states with four-particle occupations are chosen as the basis. If the initial state is prepared at the state $|0121\rangle$ as a large dipole, then only the following three Fock states $|\alpha_1\rangle = |0121\rangle$, $|\alpha_2\rangle = |0202\rangle e^{-i\omega t}$ and $|\alpha_3\rangle = |1012\rangle$ should be selected, whereas Fock states with more than $2$ particles occupied on a single site have been ignored because we only consider single photon-assisted correlated tunneling for simplicity, those Fock states violating the dipole conservation are also abandoned. Then the Floquet matrix can be approximated as $H_F = \sum_{i,j}|\alpha_i\rangle M_{ij}\langle \alpha_j|$ where $M$ is a $3\times 3$ matrix as
\begin{equation}\label{FloquetMatrix}
    M=\left(\begin{array}{ccc} \delta-\frac{8J_0^2}{U} & -2\sqrt{2}J_1 & -\frac{4J^2_0}{U} \\ -2\sqrt{2}J_1 & \frac{10J_0^2}{U} & \sqrt{2}J_1 \\ -\frac{4J^2_0}{U} & \sqrt{2}J_1 & \delta-\frac{2J_0^2}{U} \end{array}\right),
\end{equation}
where $\delta = \hbar\omega - U$ is the energy detuning and $J_0=J\mathcal{J}_0\left(\frac{A}{\hbar\omega}\right)$, $J_1=J\mathcal{J}_1\left(\frac{A}{\hbar\omega}\right)$. Here the terms $-\frac{8J_0^2}{U}=-\frac{(-2\sqrt{2}J_0)^2}{U}$ and $-\frac{2J_0^2}{U}=-\frac{(\sqrt{2}J_0)^2}{U}$ represent second-order energy shifts arising from off-resonant virtual tunneling processes mediated by the intermediate state $|0202\rangle$, which is detuned by an energy $U$. The factors $2\sqrt{2}$ and $\sqrt{2}$ come from the bosonic enhancement of the hopping matrix elements. The energy difference between the basis $|\alpha_1\rangle$ and $|\alpha_3\rangle$ comes from the finite size effect. The difference in tunneling amplitudes between $|\alpha_1\rangle \leftrightarrow |\alpha_2\rangle$ and $|\alpha_2\rangle \leftrightarrow |\alpha_3\rangle$ in Eq. (\ref{FloquetMatrix}) comes from the asymmetry of initial state, where the dipole is directed to the right along the lattice. Instead, if the initial state is chosen as $\frac{1}{\sqrt{2}}\hat{b}^\dag_j \hat{b}_{j+2} |\mathrm{MI}\rangle$ whose dipole moment is directed to the left, all our discussions should have a reflection.

\begin{figure}[htbp]
\includegraphics[width=0.9\columnwidth]{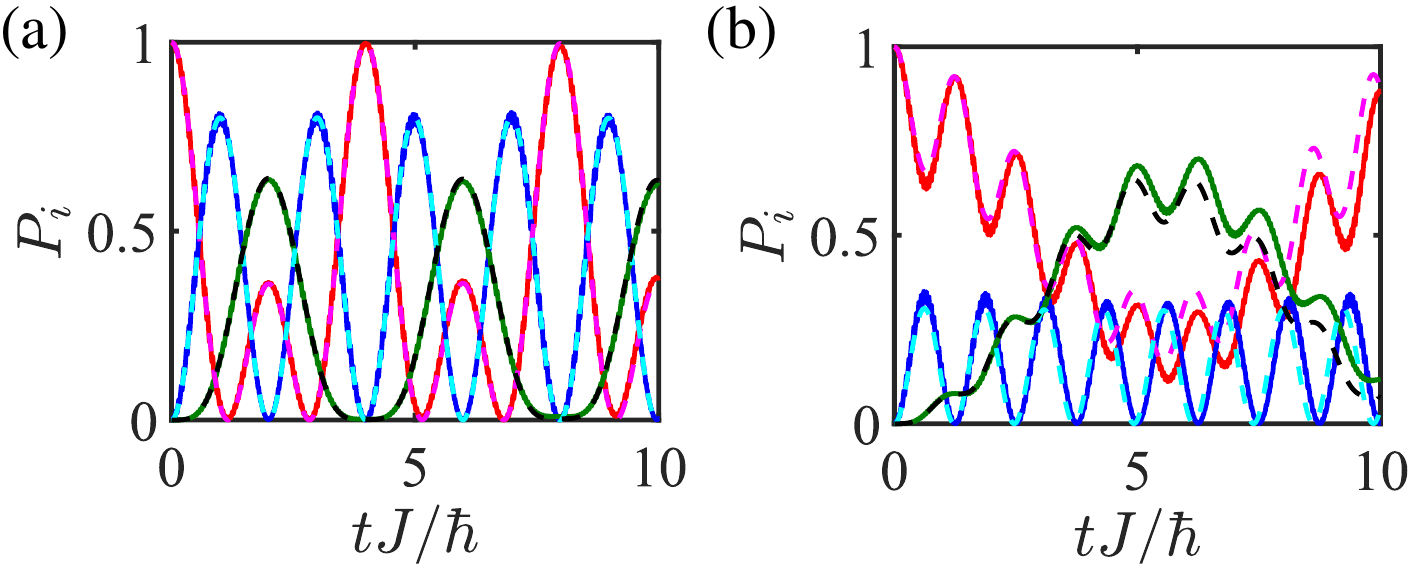}
\caption{Time evolution of the state populations $P_{i}=|\langle \psi(t)|\alpha_{i}\rangle|^2$ for the basis states $|\alpha_1\rangle$, $|\alpha_2\rangle$ and $|\alpha_3\rangle$. Solid curves depict the exact simulations generated by the Hamiltonian Eq. (\ref{fullham}): red, blue, and green correspond to $P_1$, $P_2$ and $P_3$, respectively. Dash curves show the approximate results obtained from Eq. (\ref{FloquetMatrix}) in the limit $J^2_0/U\approx 0$: magenta, cyan, and black correspond to $P_1$, $P_2$ and $P_3$. The parameters are $J = 1,  U = 120J$, and $A/\hbar\omega = 1.2$. (a) Resonant case, $\delta=0$. (b) Near-resonant regime $\delta=4J$.}
\label{fig:L4_evolution}
\end{figure}

The dynamic process is determined by the driving amplitude $A$ in the resonant case where $\delta = 0$. For small $A\sim J\ll U, \hbar\omega$, the parameters $J_0\approx J$ and $J_1\approx\frac{JA}{2\hbar\omega}$. In this regime, both $J_1$ and $J^2_0/U$ are of the same order as $J_D=4J^2/U$ which is significantly smaller than $J$. Consequently, the dynamics is almost constrained on the time scale of $\sim 1/J$. As $A$ increases, those off-diagonal terms proportional to $J_1$ become significant and no longer negligible, while the terms proportional to $J^2_0/U$ remain small and become irrelevant. If we set $J^2_0/U\approx 0$ as an approximation, Eq. (\ref{FloquetMatrix}) can be diagonalized analytically. In the resonant case $\delta=0$, the three eigenenergies are $0$ and $\mp\sqrt{10}J_1$ with the corresponding eigenstates as $\frac{1}{\sqrt{5}}(1,0,2)^\mathrm{T}$ and $\frac{1}{\sqrt{10}}(-2,\mp\sqrt{5},1)^\mathrm{T}$ respectively. Then the instantaneous state at time $t$ can be calculated as
\begin{eqnarray}\label{dynamic4site}
\begin{aligned}
    &|\psi(t)\rangle = \frac{1}{5}\Big\{\Big[4\cos(\frac{\sqrt{10}J_1t}{\hbar})+1\Big]|\alpha_1\rangle  \\
    +& 2\sqrt{5}i\sin(\frac{\sqrt{10}J_1t}{\hbar})|\alpha_2\rangle + 2\Big[1-\cos(\frac{\sqrt{10}J_1t}{\hbar})\Big]|\alpha_3\rangle\Big\}.
\end{aligned}
\end{eqnarray}
As shown in Fig.\ref{fig:L4_evolution}(a), the populations of these three states exhibit periodic oscillation with the period
\begin{equation}\label{oscillationperiod}
    T=\frac{2\pi\hbar}{\sqrt{10}J_1}.
\end{equation}

All of the above discussions are focused on the resonant case $\delta=0$. For a finite $\delta$, the oscillations are governed by two incommensurate parameters $\delta$ and $J_1$, which break the periodicity of the oscillation, as shown in Fig.\ref{fig:L4_evolution}(b). However, a finite $\delta$ does not qualitatively affect the dynamics, and the splitting and recombination of the large dipole remain observable. In Fig.\ref{fig:L4_evolution}, the simulation results obtained from the original Hamiltonian (\ref{fullham}) and the Floquet matrix (\ref{FloquetMatrix}) exhibit excellent agreement, which means that our effective model works very well.

\section{ Derivation of energy correction $\delta'$ in effective Hamiltonian (\ref{effhamdipole})}
\label{sec:more-site}
Unlike the four-site lattice, when the lattice size is large enough and the finite size effect
is ignored, the system recovers spatial translation invariance, and the dynamics includes not only the splitting of large dipoles, but also the movement of small dipoles. The estimation of tunneling amplitudes is similar to the four-site case, as shown in Eq. (\ref{effhamdipole}). The only difference is the estimation of energy shift between Fock states with one large dipole and two small dipoles.
The shift $-\frac{20J^2_0}{U}$ is a second-order energy correction generated by off-resonant virtual tunneling processes. Physically, a large dipole state  $|..1101211..\rangle$ can virtually couple to two states with neighboring small dipoles $|..1102021..\rangle$ and $|..1020211..\rangle$ with the same energy mismatch $U$ but different tunneling amplitude $-2\sqrt{2}J_0$ and $\sqrt{2}J_0$ respectively. In the bulk, two symmetry-related virtual processes contribute, yielding a shift $-(8+2)\frac{J^2_0}{U}=-\frac{10J^2_0}{U}$, the two small dipoles obtains a shift $\frac{10J^2_0}{U}$ from reverse processes, so that the total shift becomes $-\frac{20J^2_0}{U}$. Consequently, the effective detuning between Fock states with one large dipole and two small dipoles is shifted by $-\frac{20J^2_0}{U}$, leading to $\delta^\prime = \delta-\frac{20J^2_0}{U}\approx \delta$ in Eq. (\ref{effhamdipole})

\section{The mapping to hardcore bosons in a one-dimensional chain}
\label{sec:appendix}

As discussed in Sec. \ref{sec:dipole dynamics}, the dynamics of a large dipole with dipole moment $2$ can be mapped onto a two hardcore-boson system in one-dimensional chain with density-dependent tunneling and nearest-neighbor interaction. This model can be extended to more general cases.
\begin{figure}[htbp]
\includegraphics[width=1\columnwidth]{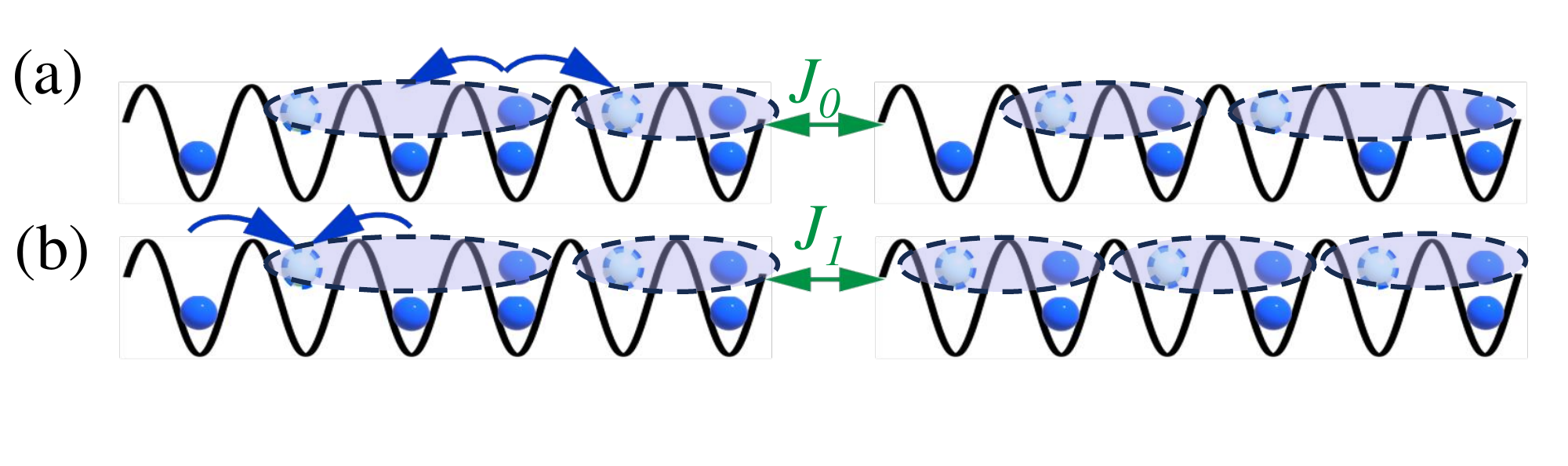}
\caption{Schemes of two distinct dipole-splitting processes. The shaded area surrounded by dashed line represents a dipole. (a) The dynamics of dipoles with conserved dipole number, which depend on static part of tensor electric field. (b) The dynamics of dipoles with non-conserved dipole number, which depend on photon-assisted correlated tunneling. The total dipole moment remains exactly conserved in both regimes.}
\label{fig:dipole number}
\end{figure}

It is straightforward to map a large dipole with dipole moment $n>2$, defined by $\hat{D}^\dag_{n,m}|\mathrm{MI}\rangle = \frac{1}{\sqrt{2}}\hat{b}_m\hat{b}^\dag_{m+n}|\mathrm{MI}\rangle$, onto a one-dimensional system of $n$ hardcore-boson where the $n$ bosons occupy the contiguous sites from $m$ to $m+n-1$, which can be defined as $|{\bf 1}_m\cdots{\bf 1}_{m+n-1}\rangle$. The splitting of this large dipole can be expressed as $2\sqrt{2}J_1|{\bf 1}_m \cdots{\bf 1}_{m+n-1}\rangle \langle {\bf 1}_m \cdots{\bf 1}_{m+n-2} {\bf 1}_{m+n}| + \sqrt{2}J_1|{\bf 1}_m \cdots{\bf 1}_{m+n-1}\rangle \langle {\bf 1}_{m-1} {\bf 1}_{m+1} \cdots{\bf 1}_{m+n-1}| + h.c.$, which corresponding to a single particle tunneling with amplitude proportional to $J_1$. As discussed in main text, all the splitting and recombination processes can be separated into two different types. One of them changes the number of dipoles by $1$, as shown in Fig. \ref{fig:dipole number} (b), which can be mapped onto the single particle tunneling as from $|{\bf 1}_m \cdots{\bf 1}_{m+n-1}\rangle$ to $|{\bf 1}_m \cdots{\bf 1}_{m+n-2} {\bf 1}_{m+n}\rangle$. These terms correspond to the photon-assisted correlated tunneling and their amplitudes depend on $J_1$. The other terms which keeping the number of dipoles as shown in Fig. \ref{fig:dipole number} (a) can be mapped to another type of single particle tunneling from $|{\bf 1}_m \cdots{\bf 1}_{m+n-2} {\bf 1}_{m+n}\rangle$ to $|{\bf 1}_m \cdots{\bf 1}_{m+n-3} {\bf 1}_{m+n-1}{\bf 1}_{m+n}\rangle$. These terms are associated with the static correlated tunneling whose amplitudes depend on $J_0$. So for the model of hardcore-boson system, the single particle tunneling can also be separated into two different parts, one of them changes the number of adjacent bosons by $1$ with amplitude depending on $J_1$, the other one preserves the number of adjacent bosons with amplitude depending on $J_0$. So this is a hardcore-boson system with density-dependent tunneling. The energy detuning can be considered as a nearest-neighbor interaction.  

\begin{figure}[htbp]
\includegraphics[width=1\columnwidth]{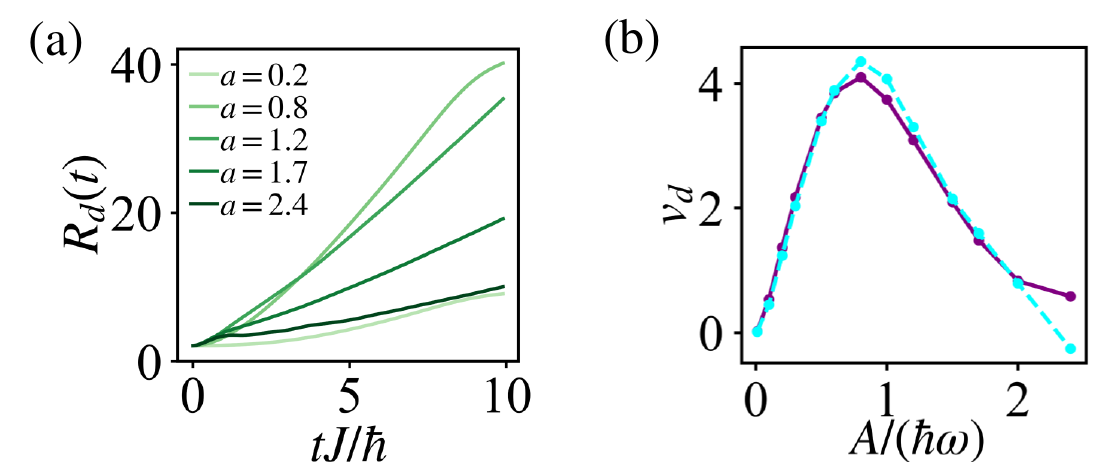}
\caption{The numerical simulation of the dynamics of resonantly driven dipole excitations in a one-dimensional lattice of length $L=61$ using TEBD. (a) The evolution of wave packet radius $R_d(t)$ of dipole excitation with moment $3$ plotted as a function of  time for various driving amplitudes $A$. (b) Expansion speed $v_d=\frac{dR_d(t)}{dt}$ for various driving amplitudes $A$, parameterized by $a=A/(\hbar\omega)$. The purple curve corresponds to a dipole excitation with dipole moment 
$3$, while the cyan curve corresponds to a dipole moment of $4$.}
\label{fig:large moment}
\end{figure}

This model still works for fracton dynamics. A single particle fracton on the background of $|\mathrm{MI}\rangle$ as $\hat{b}^\dag_m|\mathrm{MI}\rangle$ can be mapped to a one-dimensional chain with $m$ adjacent hardcore bosons occupied from site $1$ to $m$(or equivalently occupied from site $m$ to $L$). The tunneling of these hardcore bosons follows the same rules as in the above discussion. Fig. \ref{fig:large moment}(a) shows the evolution of the wave packet radius $R_d(t)$ of dipole excitation with moment $3$ and the corresponding expansion speed $v_d$ is shown in Fig. \ref{fig:large moment}(b).

\end{appendix}
\bibliography{main}

@article{thouless1982,
  title = {Quantized Hall Conductance in a Two-Dimensional Periodic Potential},
  author = {Thouless, D. J. and Kohmoto, M. and Nightingale, M. P. and den Nijs, M.},
  journal = {Phys. Rev. Lett.},
  volume = {49},
  issue = {6},
  pages = {405--408},
  numpages = {0},
  year = {1982},
  month = {Aug},
  publisher = {American Physical Society},
  doi = {10.1103/PhysRevLett.49.405},
  url = {https://link.aps.org/doi/10.1103/PhysRevLett.49.405}
}

@article{tsui1982,
  title = {Two-Dimensional Magnetotransport in the Extreme Quantum Limit},
  author = {Tsui, D. C. and Stormer, H. L. and Gossard, A. C.},
  journal = {Phys. Rev. Lett.},
  volume = {48},
  issue = {22},
  pages = {1559--1562},
  numpages = {0},
  year = {1982},
  month = {May},
  publisher = {American Physical Society},
  doi = {10.1103/PhysRevLett.48.1559},
  url = {https://link.aps.org/doi/10.1103/PhysRevLett.48.1559}
}

@article{stormer1999fractional,
  title = {The fractional quantum Hall effect},
  author = {Stormer, Horst L. and Tsui, Daniel C. and Gossard, Arthur C.},
  journal = {Rev. Mod. Phys.},
  volume = {71},
  issue = {2},
  pages = {S298--S305},
  numpages = {0},
  year = {1999},
  month = {Mar},
  publisher = {American Physical Society},
  doi = {10.1103/RevModPhys.71.S298},
  url = {https://link.aps.org/doi/10.1103/RevModPhys.71.S298}
}

@article{konig2007,
author = {Markus König  and Steffen Wiedmann  and Christoph Brüne  and Andreas Roth  and Hartmut Buhmann  and Laurens W. Molenkamp  and Xiao-Liang Qi  and Shou-Cheng Zhang },
title = {Quantum Spin Hall Insulator State in HgTe Quantum Wells},
journal = {Science},
volume = {318},
number = {5851},
pages = {766-770},
year = {2007},
doi = {10.1126/science.1148047},
URL = {https://www.science.org/doi/abs/10.1126/science.1148047},
}

@article{vijay2015,
  title = {A new kind of topological quantum order: A dimensional hierarchy of quasiparticles built from stationary excitations},
  author = {Vijay, Sagar and Haah, Jeongwan and Fu, Liang},
  journal = {Phys. Rev. B},
  volume = {92},
  issue = {23},
  pages = {235136},
  numpages = {11},
  year = {2015},
  month = {Dec},
  publisher = {American Physical Society},
  doi = {10.1103/PhysRevB.92.235136},
  url = {https://link.aps.org/doi/10.1103/PhysRevB.92.235136}
}

@article{vijay2016,
  title = {Fracton topological order, generalized lattice gauge theory, and duality},
  author = {Vijay, Sagar and Haah, Jeongwan and Fu, Liang},
  journal = {Phys. Rev. B},
  volume = {94},
  issue = {23},
  pages = {235157},
  numpages = {9},
  year = {2016},
  month = {Dec},
  publisher = {American Physical Society},
  doi = {10.1103/PhysRevB.94.235157},
  url = {https://link.aps.org/doi/10.1103/PhysRevB.94.235157}
}

@article{pretko1,
  title = {Subdimensional particle structure of higher rank $U(1)$ spin liquids},
  author = {Pretko, Michael},
  journal = {Phys. Rev. B},
  volume = {95},
  issue = {11},
  pages = {115139},
  numpages = {11},
  year = {2017},
  month = {Mar},
  publisher = {American Physical Society},
  doi = {10.1103/PhysRevB.95.115139},
  url = {https://link.aps.org/doi/10.1103/PhysRevB.95.115139}
}

@article{pretko2,
  title = {Generalized electromagnetism of subdimensional particles: A spin liquid story},
  author = {Pretko, Michael},
  journal = {Phys. Rev. B},
  volume = {96},
  issue = {3},
  pages = {035119},
  numpages = {16},
  year = {2017},
  month = {Jul},
  publisher = {American Physical Society},
  doi = {10.1103/PhysRevB.96.035119},
  url = {https://link.aps.org/doi/10.1103/PhysRevB.96.035119}
}

@article{prem2018,
  title = {Pinch point singularities of tensor spin liquids},
  author = {Prem, Abhinav and Vijay, Sagar and Chou, Yang-Zhi and Pretko, Michael and Nandkishore, Rahul M.},
  journal = {Phys. Rev. B},
  volume = {98},
  issue = {16},
  pages = {165140},
  numpages = {8},
  year = {2018},
  month = {Oct},
  publisher = {American Physical Society},
  doi = {10.1103/PhysRevB.98.165140},
  url = {https://link.aps.org/doi/10.1103/PhysRevB.98.165140}
}

@article{nandkishore2019,
   author = "Nandkishore, Rahul M. and Hermele, Michael",
   title = "Fractons", 
   journal= "Annual Review of Condensed Matter Physics",
   year = "2019",
   volume = "10",
   number = "Volume 10, 2019",
   pages = "295-313",
   doi = "https://doi.org/10.1146/annurev-conmatphys-031218-013604",
   url = "https://www.annualreviews.org/content/journals/10.1146/annurev-conmatphys-031218-013604",
   publisher = "Annual Reviews",
   issn = "1947-5462"
  }

@article{pretko2020,
author = {Pretko, Michael and Chen, Xie and You, Yizhi},
title = {Fracton phases of matter},
journal = {International Journal of Modern Physics A},
volume = {35},
number = {06},
pages = {2030003},
year = {2020},
doi = {10.1142/S0217751X20300033},
URL = { https://doi.org/10.1142/S0217751X20300033},
}

@article{ma2018fracton,
  title = {Fracton topological order from the Higgs and partial-confinement mechanisms of rank-two gauge theory},
  author = {Ma, Han and Hermele, Michael and Chen, Xie},
  journal = {Phys. Rev. B},
  volume = {98},
  issue = {3},
  pages = {035111},
  numpages = {15},
  year = {2018},
  month = {Jul},
  publisher = {American Physical Society},
  doi = {10.1103/PhysRevB.98.035111},
  url = {https://link.aps.org/doi/10.1103/PhysRevB.98.035111}
}

@article{pretko2018fracton,
  title = {Fracton-Elasticity Duality},
  author = {Pretko, Michael and Radzihovsky, Leo},
  journal = {Phys. Rev. Lett.},
  volume = {120},
  issue = {19},
  pages = {195301},
  numpages = {7},
  year = {2018},
  month = {May},
  publisher = {American Physical Society},
  doi = {10.1103/PhysRevLett.120.195301},
  url = {https://link.aps.org/doi/10.1103/PhysRevLett.120.195301}
}

@article{pretko2018fracton1,
  title = {The fracton gauge principle},
  author = {Pretko, Michael},
  journal = {Phys. Rev. B},
  volume = {98},
  issue = {11},
  pages = {115134},
  numpages = {6},
  year = {2018},
  month = {Sep},
  publisher = {American Physical Society},
  doi = {10.1103/PhysRevB.98.115134},
  url = {https://link.aps.org/doi/10.1103/PhysRevB.98.115134}
}

@article{prem2017glassy,
  title = {Glassy quantum dynamics in translation invariant fracton models},
  author = {Prem, Abhinav and Haah, Jeongwan and Nandkishore, Rahul},
  journal = {Phys. Rev. B},
  volume = {95},
  issue = {15},
  pages = {155133},
  numpages = {14},
  year = {2017},
  month = {Apr},
  publisher = {American Physical Society},
  doi = {10.1103/PhysRevB.95.155133},
  url = {https://link.aps.org/doi/10.1103/PhysRevB.95.155133}
}

@article{haah2011local,
  title = {Local stabilizer codes in three dimensions without string logical operators},
  author = {Haah, Jeongwan},
  journal = {Phys. Rev. A},
  volume = {83},
  issue = {4},
  pages = {042330},
  numpages = {16},
  year = {2011},
  month = {Apr},
  publisher = {American Physical Society},
  doi = {10.1103/PhysRevA.83.042330},
  url = {https://link.aps.org/doi/10.1103/PhysRevA.83.042330}
}

@article{ma2017fracton,
  title = {Fracton topological order via coupled layers},
  author = {Ma, Han and Lake, Ethan and Chen, Xie and Hermele, Michael},
  journal = {Phys. Rev. B},
  volume = {95},
  issue = {24},
  pages = {245126},
  numpages = {18},
  year = {2017},
  month = {Jun},
  publisher = {American Physical Society},
  doi = {10.1103/PhysRevB.95.245126},
  url = {https://link.aps.org/doi/10.1103/PhysRevB.95.245126}
}

@article{pai2019localization,
  title = {Localization in Fractonic Random Circuits},
  author = {Pai, Shriya and Pretko, Michael and Nandkishore, Rahul M.},
  journal = {Phys. Rev. X},
  volume = {9},
  issue = {2},
  pages = {021003},
  numpages = {21},
  year = {2019},
  month = {Apr},
  publisher = {American Physical Society},
  doi = {10.1103/PhysRevX.9.021003},
  url = {https://link.aps.org/doi/10.1103/PhysRevX.9.021003}
}

@article{myerson2022construction,
  title = {Construction of Fractal Order and Phase Transition with Rydberg Atoms},
  author = {Myerson-Jain, Nayan E. and Yan, Stephen and Weld, David and Xu, Cenke},
  journal = {Phys. Rev. Lett.},
  volume = {128},
  issue = {1},
  pages = {017601},
  numpages = {6},
  year = {2022},
  month = {Jan},
  publisher = {American Physical Society},
  doi = {10.1103/PhysRevLett.128.017601},
  url = {https://link.aps.org/doi/10.1103/PhysRevLett.128.017601}
}

@article{yang2020,
  title={Observation of gauge invariance in a 71-site Bose--Hubbard quantum simulator},
  author={Yang, Bing and Sun, Hui and Ott, Robert and Wang, Han-Yi and Zache, Torsten V and Halimeh, Jad C and Yuan, Zhen-Sheng and Hauke, Philipp and Pan, Jian-Wei},
  journal={Nature},
  volume={587},
  number={7834},
  pages={392--396},
  year={2020},
  publisher={Nature Publishing Group UK London},
  doi = {https://doi.org/10.1038/s41586-020-2910-8},
  url = {https://doi.org/10.1038/s41586-020-2910-8}
}

@article{zhou2022,
author = {Zhao-Yu Zhou  and Guo-Xian Su  and Jad C. Halimeh  and Robert Ott  and Hui Sun  and Philipp Hauke  and Bing Yang  and Zhen-Sheng Yuan  and Jürgen Berges  and Jian-Wei Pan },
title = {Thermalization dynamics of a gauge theory on a quantum simulator},
journal = {Science},
volume = {377},
number = {6603},
pages = {311-314},
year = {2022},
doi = {10.1126/science.abl6277},
URL = {https://www.science.org/doi/abs/10.1126/science.abl6277}}

@article{scherg2021,
  title={Observing non-ergodicity due to kinetic constraints in tilted Fermi-Hubbard chains},
  author={Scherg, Sebastian and Kohlert, Thomas and Sala, Pablo and Pollmann, Frank and Hebbe Madhusudhana, Bharath and Bloch, Immanuel and Aidelsburger, Monika},
  journal={Nature Communications},
  volume={12},
  number={1},
  pages={4490},
  year={2021},
  publisher={Nature Publishing Group UK London},
  doi = {https://doi.org/10.1038/s41467-021-24726-0},
  url = {https://doi.org/10.1038/s41467-021-24726-0}
}

@article{kohlert2023,
  title = {Exploring the Regime of Fragmentation in Strongly Tilted Fermi-Hubbard Chains},
  author = {Kohlert, Thomas and Scherg, Sebastian and Sala, Pablo and Pollmann, Frank and Hebbe Madhusudhana, Bharath and Bloch, Immanuel and Aidelsburger, Monika},
  journal = {Phys. Rev. Lett.},
  volume = {130},
  issue = {1},
  pages = {010201},
  numpages = {7},
  year = {2023},
  month = {Jan},
  publisher = {American Physical Society},
  doi = {10.1103/PhysRevLett.130.010201},
  url = {https://link.aps.org/doi/10.1103/PhysRevLett.130.010201}
}

@article{adler2024,
  title={Observation of Hilbert space fragmentation and fractonic excitations in 2D},
  author={Adler, Daniel and Wei, David and Will, Melissa and Srakaew, Kritsana and Agrawal, Suchita and Weckesser, Pascal and Moessner, Roderich and Pollmann, Frank and Bloch, Immanuel and Zeiher, Johannes},
  journal={Nature},
  pages={1--6},
  year={2024},
  publisher={Nature Publishing Group UK London},
  doi = {https://doi.org/10.1038/s41586-024-08188-0},
  url = {https://doi.org/10.1038/s41586-024-08188-0}
}

@article{will2024,
  title = {Realization of Hilbert Space Fragmentation and Fracton Dynamics in Two Dimensions},
  author = {Will, Melissa and Moessner, Roderich and Pollmann, Frank},
  journal = {Phys. Rev. Lett.},
  volume = {133},
  issue = {19},
  pages = {196301},
  numpages = {6},
  year = {2024},
  month = {Nov},
  publisher = {American Physical Society},
  doi = {10.1103/PhysRevLett.133.196301},
  url = {https://link.aps.org/doi/10.1103/PhysRevLett.133.196301}
}

@article{lake2022,
  title = {Dipolar Bose-Hubbard model},
  author = {Lake, Ethan and Hermele, Michael and Senthil, T.},
  journal = {Phys. Rev. B},
  volume = {106},
  issue = {6},
  pages = {064511},
  numpages = {14},
  year = {2022},
  month = {Aug},
  publisher = {American Physical Society},
  doi = {10.1103/PhysRevB.106.064511},
  url = {https://link.aps.org/doi/10.1103/PhysRevB.106.064511}
}

@article{lake2023,
  title = {Dipole condensates in tilted Bose-Hubbard chains},
  author = {Lake, Ethan and Lee, Hyun-Yong and Han, Jung Hoon and Senthil, T.},
  journal = {Phys. Rev. B},
  volume = {107},
  issue = {19},
  pages = {195132},
  numpages = {20},
  year = {2023},
  month = {May},
  publisher = {American Physical Society},
  doi = {10.1103/PhysRevB.107.195132},
  url = {https://link.aps.org/doi/10.1103/PhysRevB.107.195132}
}

@article{Zechmann2023,
  title = {Fractonic Luttinger liquids and supersolids in a constrained Bose-Hubbard model},
  author = {Zechmann, Philip and Altman, Ehud and Knap, Michael and Feldmeier, Johannes},
  journal = {Phys. Rev. B},
  volume = {107},
  issue = {19},
  pages = {195131},
  numpages = {16},
  year = {2023},
  month = {May},
  publisher = {American Physical Society},
  doi = {10.1103/PhysRevB.107.195131},
  url = {https://link.aps.org/doi/10.1103/PhysRevB.107.195131}
}

@article{Zechmann2024,
  title = {Dynamical spectral response of fractonic quantum matter},
  author = {Zechmann, Philip and Boesl, Julian and Feldmeier, Johannes and Knap, Michael},
  journal = {Phys. Rev. B},
  volume = {109},
  issue = {12},
  pages = {125137},
  numpages = {10},
  year = {2024},
  month = {Mar},
  publisher = {American Physical Society},
  doi = {10.1103/PhysRevB.109.125137},
  url = {https://link.aps.org/doi/10.1103/PhysRevB.109.125137}
}

@article{xu2024,
  title={Multipolar condensates and multipolar Josephson effects},
  author={Xu, Wenhui and Lv, Chenwei and Zhou, Qi},
  journal={Nature Communications},
  volume={15},
  number={1},
  pages={4786},
  year={2024},
  publisher={Nature Publishing Group UK London},
  doi = {https://doi.org/10.1038/s41467-024-48907-9},
  url = {https://doi.org/10.1038/s41467-024-48907-9}
}

@article{boesl2024,
  title = {Deconfinement Dynamics of Fractons in Tilted Bose-Hubbard Chains},
  author = {Boesl, Julian and Zechmann, Philip and Feldmeier, Johannes and Knap, Michael},
  journal = {Phys. Rev. Lett.},
  volume = {132},
  issue = {14},
  pages = {143401},
  numpages = {7},
  year = {2024},
  month = {Apr},
  publisher = {American Physical Society},
  doi = {10.1103/PhysRevLett.132.143401},
  url = {https://link.aps.org/doi/10.1103/PhysRevLett.132.143401}
}

@article{zhang2025synthetic,
  title = {Synthetic tensor gauge fields},
  author = {Zhang, Shaoliang and Lv, Chenwei and Zhou, Qi},
  journal = {Phys. Rev. Res.},
  volume = {7},
  issue = {1},
  pages = {013013},
  numpages = {10},
  year = {2025},
  month = {Jan},
  publisher = {American Physical Society},
  doi = {10.1103/PhysRevResearch.7.013013},
  url = {https://link.aps.org/doi/10.1103/PhysRevResearch.7.013013}
}

@article{moudgalya2022quantum,
doi = {10.1088/1361-6633/ac73a0},
url = {https://dx.doi.org/10.1088/1361-6633/ac73a0},
year = {2022},
month = {jul},
publisher = {IOP Publishing},
volume = {85},
number = {8},
pages = {086501},
author = {Moudgalya, Sanjay and Bernevig, B Andrei and Regnault, Nicolas},
title = {Quantum many-body scars and Hilbert space fragmentation: a review of exact results},
journal = {Reports on Progress in Physics}
}

@article{sala2020ergodicity,
  title = {Ergodicity Breaking Arising from Hilbert Space Fragmentation in Dipole-Conserving Hamiltonians},
  author = {Sala, Pablo and Rakovszky, Tibor and Verresen, Ruben and Knap, Michael and Pollmann, Frank},
  journal = {Phys. Rev. X},
  volume = {10},
  issue = {1},
  pages = {011047},
  numpages = {19},
  year = {2020},
  month = {Feb},
  publisher = {American Physical Society},
  doi = {10.1103/PhysRevX.10.011047},
  url = {https://link.aps.org/doi/10.1103/PhysRevX.10.011047}
}

@article{khemani2020localization,
  title = {Localization from Hilbert space shattering: From theory to physical realizations},
  author = {Khemani, Vedika and Hermele, Michael and Nandkishore, Rahul},
  journal = {Phys. Rev. B},
  volume = {101},
  issue = {17},
  pages = {174204},
  numpages = {17},
  year = {2020},
  month = {May},
  publisher = {American Physical Society},
  doi = {10.1103/PhysRevB.101.174204},
  url = {https://link.aps.org/doi/10.1103/PhysRevB.101.174204}
}

@article{chin2010feshbach,
  title = {Feshbach resonances in ultracold gases},
  author = {Chin, Cheng and Grimm, Rudolf and Julienne, Paul and Tiesinga, Eite},
  journal = {Rev. Mod. Phys.},
  volume = {82},
  issue = {2},
  pages = {1225--1286},
  numpages = {0},
  year = {2010},
  month = {Apr},
  publisher = {American Physical Society},
  doi = {10.1103/RevModPhys.82.1225},
  url = {https://link.aps.org/doi/10.1103/RevModPhys.82.1225}
}

@article{eckardt2015high,
doi = {10.1088/1367-2630/17/9/093039},
url = {https://dx.doi.org/10.1088/1367-2630/17/9/093039},
year = {2015},
month = {sep},
publisher = {IOP Publishing},
volume = {17},
number = {9},
pages = {093039},
author = {Eckardt, André and Anisimovas, Egidijus},
title = {High-frequency approximation for periodically driven quantum systems from a Floquet-space perspective},
journal = {New Journal of Physics}}

@article{goldman2014periodically,
  title = {Periodically Driven Quantum Systems: Effective Hamiltonians and Engineered Gauge Fields},
  author = {Goldman, N. and Dalibard, J.},
  journal = {Phys. Rev. X},
  volume = {4},
  issue = {3},
  pages = {031027},
  numpages = {29},
  year = {2014},
  month = {Aug},
  publisher = {American Physical Society},
  doi = {10.1103/PhysRevX.4.031027},
  url = {https://link.aps.org/doi/10.1103/PhysRevX.4.031027}
}

@article{rahav2003effective,
  title = {Effective Hamiltonians for periodically driven systems},
  author = {Rahav, Saar and Gilary, Ido and Fishman, Shmuel},
  journal = {Phys. Rev. A},
  volume = {68},
  issue = {1},
  pages = {013820},
  numpages = {18},
  year = {2003},
  month = {Jul},
  publisher = {American Physical Society},
  doi = {10.1103/PhysRevA.68.013820},
  url = {https://link.aps.org/doi/10.1103/PhysRevA.68.013820}
}

@article{sias2008observation,
  title = {Observation of Photon-Assisted Tunneling in Optical Lattices},
  author = {Sias, C. and Lignier, H. and Singh, Y. P. and Zenesini, A. and Ciampini, D. and Morsch, O. and Arimondo, E.},
  journal = {Phys. Rev. Lett.},
  volume = {100},
  issue = {4},
  pages = {040404},
  numpages = {4},
  year = {2008},
  month = {Feb},
  publisher = {American Physical Society},
  doi = {10.1103/PhysRevLett.100.040404},
  url = {https://link.aps.org/doi/10.1103/PhysRevLett.100.040404}
}

@Article{hauschild2018efficient,
	title={{Efficient numerical simulations with Tensor Networks: Tensor Network Python (TeNPy)}},
	author={Johannes Hauschild and Frank Pollmann},
	journal={SciPost Phys. Lect. Notes},
	pages={5},
	year={2018},
	publisher={SciPost},
	doi={10.21468/SciPostPhysLectNotes.5},
	url={https://scipost.org/10.21468/SciPostPhysLectNotes.5},
}

@article{Martin,
  title = {Topological Frequency Conversion in Strongly Driven Quantum Systems},
  author = {Martin, Ivar and Refael, Gil and Halperin, Bertrand},
  journal = {Phys. Rev. X},
  volume = {7},
  issue = {4},
  pages = {041008},
  numpages = {20},
  year = {2017},
  month = {Oct},
  publisher = {American Physical Society},
  doi = {10.1103/PhysRevX.7.041008},
  url = {https://link.aps.org/doi/10.1103/PhysRevX.7.041008}
}

@article{Nathan,
  title = {Quasiperiodic Floquet-Thouless Energy Pump},
  author = {Nathan, Frederik and Ge, Rongchun and Gazit, Snir and Rudner, Mark and Kolodrubetz, Michael},
  journal = {Phys. Rev. Lett.},
  volume = {127},
  issue = {16},
  pages = {166804},
  numpages = {6},
  year = {2021},
  month = {Oct},
  publisher = {American Physical Society},
  doi = {10.1103/PhysRevLett.127.166804},
  url = {https://link.aps.org/doi/10.1103/PhysRevLett.127.166804}
}

@article{Weitenberg2021,
  title     = {Tailoring quantum gases by Floquet engineering},
  author    = {Weitenberg, Christof and Simonet, Juliette},
  journal   = {Nature Physics},
  volume    = {17},
  issue     = {12},
  pages     = {1342--1348},
  numpages  = {7},
  year      = {2021},
  month     = {Aug},
  publisher = {Springer Nature},
  doi       = {10.1038/s41567-021-01316-x},
  url       = {https://www.nature.com/articles/s41567-021-01316-x}
}

@article{Theis,
  title = {Tuning the Scattering Length with an Optically Induced Feshbach Resonance},
  author = {Theis, M. and Thalhammer, G. and Winkler, K. and Hellwig, M. and Ruff, G. and Grimm, R. and Denschlag, J. Hecker},
  journal = {Phys. Rev. Lett.},
  volume = {93},
  issue = {12},
  pages = {123001},
  numpages = {4},
  year = {2004},
  month = {Sep},
  publisher = {American Physical Society},
  doi = {10.1103/PhysRevLett.93.123001},
  url = {https://link.aps.org/doi/10.1103/PhysRevLett.93.123001}
}

@article{Pai,
  title = {Dynamical Scar States in Driven Fracton Systems},
  author = {Pai, Shriya and Pretko, Michael},
  journal = {Phys. Rev. Lett.},
  volume = {123},
  issue = {13},
  pages = {136401},
  numpages = {5},
  year = {2019},
  month = {Sep},
  publisher = {American Physical Society},
  doi = {10.1103/PhysRevLett.123.136401},
  url = {https://link.aps.org/doi/10.1103/PhysRevLett.123.136401}
}

@article{Bloch2013,
  title = {Realization of the Hofstadter Hamiltonian with Ultracold Atoms in Optical Lattices},
  author = {Aidelsburger, M. and Atala, M. and Lohse, M. and Barreiro, J. T. and Paredes, B. and Bloch, I.},
  journal = {Phys. Rev. Lett.},
  volume = {111},
  issue = {18},
  pages = {185301},
  numpages = {5},
  year = {2013},
  month = {Oct},
  publisher = {American Physical Society},
  doi = {10.1103/PhysRevLett.111.185301},
  url = {https://link.aps.org/doi/10.1103/PhysRevLett.111.185301}
}

@article{Bloch2008,
  title = {Many-body physics with ultracold gases},
  author = {Bloch, Immanuel and Dalibard, Jean and Zwerger, Wilhelm},
  journal = {Rev. Mod. Phys.},
  volume = {80},
  issue = {3},
  pages = {885--964},
  numpages = {0},
  year = {2008},
  month = {Jul},
  publisher = {American Physical Society},
  doi = {10.1103/RevModPhys.80.885},
  url = {https://link.aps.org/doi/10.1103/RevModPhys.80.885}
}

@article{Jiang2023,
  title = {Kapitza Trap for Ultracold Atoms},
  author = {Jiang, Jian and Bernhart, Erik and R\"ohrle, Marvin and Benary, Jens and Beck, Marvin and Baals, Christian and Ott, Herwig},
  journal = {Phys. Rev. Lett.},
  volume = {131},
  issue = {3},
  pages = {033401},
  numpages = {6},
  year = {2023},
  month = {Jul},
  publisher = {American Physical Society},
  doi = {10.1103/PhysRevLett.131.033401},
  url = {https://link.aps.org/doi/10.1103/PhysRevLett.131.033401}
}

@article{gross2017,
  title={Quantum simulations with ultracold atoms in optical lattices},
  author={Gross, Christian and Bloch, Immanuel},
  journal={Science},
  volume={357},
  number={6355},
  pages={995--1001},
  year={2017},
  publisher={American Association for the Advancement of Science},
  doi={https://doi.org/10.1126/science.aal3837},
  url={https://doi.org/10.1126/science.aal3837}
}

@article{bakr2009,
  title={A quantum gas microscope for detecting single atoms in a Hubbard--regime optical lattice},
  author={Bakr, Waseem S and Gillen, Jonathon I and Peng, Amy and Fölling, Simon and Greiner, Markus},
  journal={Nature},
  volume={462},
  number={7269},
  pages={74--77},
  year={2009},
  publisher={Nature Publishing Group UK London},
  doi={https://doi.org/10.1038/nature08482},
  url={https://doi.org/10.1038/nature08482}
}

@article{sherson2010,
  title={Single-atom-resolved fluorescence imaging of an atomic Mott insulator},
  author={Sherson, Jacob F and Weitenberg, Christof and Endres, Manuel and Cheneau, Marc and Bloch, Immanuel and Kuhr, Stefan},
  journal={Nature},
  volume={467},
  number={7311},
  pages={68--72},
  year={2010},
  publisher={Nature Publishing Group UK London},
  doi={https://doi.org/10.1038/nature09378},
  url={https://doi.org/10.1038/nature09378}
}

@article{kim2025Multi,
  title={Multi-particle quantum walks in a dipole-conserving Bose-Hubbard model},
  author={Kim, Sooshin and Kang, Byungmin and Segura, Perrin and Li, Yanfei and Lake, Ethan and Bakkali-Hassani, Brice and Greiner, Markus},
  journal={arXiv:2511.02343},
  year={2025},
  url = {https://arxiv.org/abs/2511.02343},
}

@article{You2021,
  title = {Multipolar topological field theories: Bridging higher order topological insulators and fractons},
  author = {You, Yizhi and Burnell, F. J. and Hughes, Taylor L.},
  journal = {Phys. Rev. B},
  volume = {103},
  issue = {24},
  pages = {245128},
  numpages = {28},
  year = {2021},
  month = {Jun},
  publisher = {American Physical Society},
  doi = {10.1103/PhysRevB.103.245128},
  url = {https://link.aps.org/doi/10.1103/PhysRevB.103.245128}
}

@article{Schulz2019,
  title = {Stark Many-Body Localization},
  author = {Schulz, M. and Hooley, C. A. and Moessner, R. and Pollmann, F.},
  journal = {Phys. Rev. Lett.},
  volume = {122},
  issue = {4},
  pages = {040606},
  numpages = {5},
  year = {2019},
  month = {Jan},
  publisher = {American Physical Society},
  doi = {10.1103/PhysRevLett.122.040606},
  url = {https://link.aps.org/doi/10.1103/PhysRevLett.122.040606}
}

@article{Taylor2020,
  title = {Experimental probes of Stark many-body localization},
  author = {Taylor, S. R. and Schulz, M. and Pollmann, F. and Moessner, R.},
  journal = {Phys. Rev. B},
  volume = {102},
  issue = {5},
  pages = {054206},
  numpages = {10},
  year = {2020},
  month = {Aug},
  publisher = {American Physical Society},
  doi = {10.1103/PhysRevB.102.054206},
  url = {https://link.aps.org/doi/10.1103/PhysRevB.102.054206}
}

@article{Berislav2022,
  title = {Dynamical l-bits and persistent oscillations in Stark many-body localization},
  author = {Gunawardana, Thivan M. and Bu\ifmmode \check{c}\else \v{c}\fi{}a, Berislav},
  journal = {Phys. Rev. B},
  volume = {106},
  issue = {16},
  pages = {L161111},
  numpages = {6},
  year = {2022},
  month = {Oct},
  publisher = {American Physical Society},
  doi = {10.1103/PhysRevB.106.L161111},
  url = {https://link.aps.org/doi/10.1103/PhysRevB.106.L161111}
}

@article{Berislav2024,
  title={Protecting coherence from the environment via Stark many-body localization in a Quantum-Dot Simulator},
  author={Sarkar, Subhajit and Bu{\v{c}}a, Berislav},
  journal={Quantum},
  volume={8},
  pages={1392},
  year={2024},
  publisher={Verein zur F{\"o}rderung des Open Access Publizierens in den Quantenwissenschaften},
  url = {https://quantum-journal.org/papers/q-2024-07-02-1392/}
}

@misc{Zhang2025CompAppendix,
author       = {Zhang, Jiali and Zhang, Shaoliang},
title        = {Computational appendix of {Resonant dynamics of 1D dipole-conserving Bose-Hubbard model with time-dependent tensor electric fields}},
year         = {2025},
howpublished = {GitHub repository},
url          = {https://github.com/Hagendasis/Resonant_dynamics_of_1D_dipole-conserving_Bose-Hubbard_model}
}
\end{document}